\documentclass{article}
\usepackage{amsmath,amsfonts,setspace,cancel,url,hyperref,verbatim,soul,cite}
\makeatletter
\def\maketag@@@#1{\hbox{\m@th\normalfont\normalsize#1}}
\makeatother
\usepackage[a4paper]{geometry}
\newcommand{\be}{\begin{equation}}
\newcommand{\ee}{\end{equation}}
\newcommand{\gM}{\mathcal{H}}

\numberwithin{equation}{section}
\begin{document}

\begin{titlepage}%1
\begin{center}

\hfill  

\vskip 1.5cm

{\Large \bf Doubled strings, negative strings and null waves}

\vskip 1cm

{\bf Chris D. A. Blair} \\

\vskip 25pt

{\em Theoretische Natuurkunde, Vrije Universiteit Brussel, and the International Solvay Institutes,\\
Pleinlaan 2, B-1050 Brussels, Belgium
\vskip 5pt }

{email: {\tt cblair@vub.ac.be}} \\

\end{center}

\vskip 0.5cm

\begin{center} {\bf ABSTRACT}\\[3ex]
\end{center}

\noindent 
We revisit the fundamental string (F1) solution in the doubled formalism. We show that the wave-like solution of double field theory (DFT) corresponding to the F1/pp-wave duality pair is more properly a solution of the DFT action coupled to a doubled sigma model action. The doubled string configuration which sources the pp-wave can be thought of as static gauge with the string oriented in a dual direction. 
We also discuss the DFT solution corresponding to a vibrating string, carrying both winding and momentum.  
We further show that the solution dual to the F1 in both time and space can be viewed as a ``negative string'' solution. 
Negative branes are closely connected to certain exotic string theories which involve unusual signatures for both spacetime and brane worldvolumes. 
In order to better understand this from the doubled point of view, we construct a variant of DFT suitable for describing theories in which the fundamental string has a Euclidean worldsheet, for which T-dualities appear to change the spacetime signature. 

\end{titlepage}

\newpage
\setcounter{page}{1}
%\tableofcontents

\section{Introduction}

The goal of so-called ``doubled formalisms'' is to pair the string coordinates with their T-duals in order to achieve duality invariance as a manifest symmetry. This has been pioneered in particular by \cite{Duff:1989tf, Tseytlin:1990nb,Tseytlin:1990va,Siegel:1993th, Siegel:1993xq, Hull:2004in,Hull:2006va}, leading to doubled worldsheet actions, and has been applied to supergravity to give double field theory (DFT) \cite{Siegel:1993th, Siegel:1993xq,Hull:2009mi, Hull:2009zb, Hohm:2010jy, Hohm:2010pp}.

Solutions of DFT have been considered in \cite{Berkeley:2014nza, Berman:2014jsa, Berman:2014hna}. In \cite{Berkeley:2014nza} the DFT configuration corresponding to the T-dual pair of a fundamental string and a pp-wave was written down: this solution was shown to take the form of a null wave in the doubled spacetime. One has to make a choice as to which half of the directions of the doubled spacetime are to be considered physical. This is called a choice of ``section''. If this doubled wave is oriented in the physical section, then the solution reduces in spacetime to that of the pp-wave, while if instead it is oriented in a \emph{dual} direction then the solution appears as a string. This wavelike interpretation was supported by the calculation of the charges of this solution in \cite{Blair:2015eba, Park:2015bza,Naseer:2015fba}, from which one sees that the string winding charge/pp-wave momentum correspond simply to the conserved charge associated to translational invariance in the doubled space.

In this paper, we wish to revisit this solution. In particular, we want to clarify the nature of its source. One can construct wave-like solutions to double field theory, as to supergravity, which are specified by a particular harmonic function. However, the fundamental string (F1) solution is a solution not to the pure supergravity equations of motion, but to the action given by coupling the bulk supergravity action to a string worldsheet action \cite{Dabholkar:1989jt, Dabholkar:1990yf}. 

We shall see that the appropriate source for the doubled wave solution is indeed a doubled string worldsheet action. Intuitively, the doubled string sources a string when the worldvolume direction lies in the physical spacetime: it then can also act as a source for the pp-wave when we orient it in a dual direction. We write down the equations of motion of the full action including the source in section \ref{action}, and then show how this is solved by the doubled F1 solution in section \ref{solution}.

The doubled worldsheet action that we will use will be essentially that given in \cite{Blair:2013noa}, which is closely related to the well-known Tseytlin action \cite{Tseytlin:1990nb,Tseytlin:1990va}, as also considered with DFT applications in mind in \cite{Copland:2011wx}. The action of \cite{Blair:2013noa} is inspired by the Hamiltonian form of the string worldsheet action, and so in particular retains the Virasoro constraints, which our source configuration must solve. (We will also show in appendix \ref{appws} that the doubled action due to Hull \cite{Hull:2004in,Hull:2006va} provides a possible source: presumably, the same will be true for any doubled worldsheet action which reduces to the conventional string sigma model e.g. \cite{Lee:2013hma}.
For a review of doubled worldsheet approaches, see \cite{Berman:2013eva}.)

This allows us to complete a cycle of ideas connecting the doubled worldsheet with the doubled string. The equations of motion of double field theory arise as beta-functional equations for doubled string actions \cite{Berman:2007xn,Berman:2007yf, Copland:2011wx}: these equations have a solution which is sourced by the doubled string, and the fluctuations about this source reproduce again the self-duality equations of the doubled string \cite{Berkeley:2014nza}.

Having investigated the solution which represents a single string or wave, the next question to wonder about is that of superpositions of such solutions, for example if we want to describe a string carrying both momentum and winding in a single direction. 

It has long been established that the naive superposition of the F1 and pp-wave solution does not give a valid string theory background: the reason for this is that this configuration cannot be sourced by the usual string sigma model action. However, one can construct a solution corresponding to the background resulting from a macroscopic string carrying solely left-moving oscillations \cite{Dabholkar:1995nc, Callan:1995hn}.
This solution winds many times around one direction and carries momentum along it. It is not localised in the transverse directions, but rather traces out some non-trivial curve. The intuition is that the string has no longitudinal oscillation modes, and so if made to carry momentum in a worldvolume direction must therefore extend in the transverse directions.

In section \ref{vib}, we will embed this configuration into double field theory, using the doubled string as a source. We shall see in particular that the source configuration can also be viewed as vibrating simultaneously in the dual transverse directions. In the conventional spacetime picture, one can think of these directions as being smeared over, however in double field theory this may not be necessary. We leave further exploration of the properties of this solution for future study.  

In doubling all coordinates, one also doubles time. The doubled string solution therefore also contains the configuration which is dual to the F1 solution along both temporal and spatial worldvolume directions. This has been considered before in \cite{Sakatani:2014hba , Blair:2015eba, Park:2015bza}, where it has been noticed that this solution may be thought of as the electromagnetic dual to the exotic $5_2^2$ brane (and so is electrically charged under a bivector field which can be used alternatively to the $B$-field) and also seemingly has negative ADM mass. 

We shall show in this paper, in section \ref{Tsolution}, that this solution is in fact a ``negative string.'' 
This allows us to connect to the recent exploration in \cite{Dijkgraaf:2016lym} of ``negative branes.'' It has been argued there that negative branes are in fact the standard branes of various exotic string theories and M-theories, which were originally studied by Hull \cite{Hull:1998vg, Hull:1998ym}, using timelike dualities. These exotic theories both have unusual spacetime signatures, and contain branes whose worldvolume theories have non-standard signatures. In some cases, they are related by dualities which appear to change the signature of spacetime. 

It is likely that in order to fully understand DFT, we will need to confront the appearance of an extra time coordinate. This is one motivation for using it to explore these phenomena. Another motivation is to see if an application of DFT is to provide a framework in which to understand the theories presented in \cite{Hull:1998vg, Hull:1998ym, Dijkgraaf:2016lym} (as has already been seen for certain modified type II theories in \cite{Hohm:2011dv}). 

Here, as we mentioned, we will see firstly
 that the DFT string solution naturally includes the negative F1 solution, and discuss some of its properties as seen in the doubled formalism. We will then construct, in section \ref{EF1}, a novel variant of DFT, which applies to theories where the fundamental string has a Euclidean worldsheet. This new version of DFT, which we call DFT$^-$, differs in a particular modification of the generalised metric. 

We will also discuss, in section \ref{Esig} and appendix \ref{genlor}, the nature of signature changing duality transformations in DFT$^-$. The results of \cite{Malek:2013sp} imply that, for the groups $\mathrm{SL}(3)\times \mathrm{SL}(2)$ and $\mathrm{SL}(5)$, timelike U-dualities cannot change the signature: rather, they will generically imply the necessary inclusion of antisymmetric vector fields. We shall see that the situation in DFT$^-$ is perhaps a little different, but not without its own subtleties. 

Finally, we provide some discussion in section \ref{conclusions}. The two appendices give additional results related to the doubled worldsheet (in appendix \ref{appws}) and to DFT (in appendix \ref{appst}). 

\section{Action for DFT and doubled worldsheet} 
\label{action}

We will consider the action $S = S_{DFT} + S_{DWS}$, which describes the dynamics of the (NSNS sector) double field theory fields, the generalised metric, $\gM_{MN}$, and generalised dilaton, $d$, sourced by a doubled sigma model. 
The double field theory action is \cite{Hohm:2010pp}
\be
S_{DFT} = \frac{1}{16\pi G_{DFT}} \int d^{2D} x e^{-2d} \mathcal{R} \,,
\label{SDFT}
\ee
where the generalised Ricci scalar is given by
\be
\begin{split}
\mathcal{R} & = 4 \gM^{MN}\partial_{M}\partial_{N}d
-\partial_{M}\partial_{N}\gM^{MN} 
-4\gM^{MN}\partial_{M}d\,\partial_{N}d
+ 4 \partial_M \gM^{MN}  \partial_Nd\\
&+\frac{1}{8}\gM^{MN}\partial_{M}\gM^{KL}
\partial_{N}\gM_{KL}-\frac{1}{2}\gM^{MN}\partial_{M}\gM^{KL}
\partial_{K}\gM_{NL}\,.
\end{split}
\label{RDFT}
\ee
We denote the doubled coordinates by $x^M = ( x^i, \tilde x_i)$. 

The generalised dilaton, $d$, is a T-duality invariant, while the generalised metric, $\gM_{MN}$, is a rank 2 tensor under $O(D,D)$ transformations. It is symmetric and constrained to satisfy $\gM_{MP} \eta^{PQ} \gM_{QN} = \eta_{MN}$, where 
\be
\eta_{MN} = \begin{pmatrix} 0 & \delta_i{}^j \\ \delta^i{}_j & 0 \end{pmatrix} \,,
\label{eta} 
\ee
is the defining $O(D,D)$ structure. We use this to raise and lower indices, so that the inverse of $\gM_{MN}$ is $\gM^{MN} = \eta^{MP} \eta^{NQ} \gM_{PQ}$. These conditions mean that $\gM_{MN}$ parametrises the coset $O(D,D) / O(1,D-1) \times O(1,D-1)$ (very frequently one ignores time, in which case the coset is $O(D,D)/O(D)\times O(D)$). This denominator group $O(1,D-1) \times O(1,D-1)$ is the generalised Lorentz group. 

The other local transformations of DFT are generalised diffeomorphisms, under which $\gM_{MN}$ is a rank 2 tensor and $e^{-2d}$ has weight one,
\be
\delta_\Lambda \gM_{MN} = \Lambda^P \partial_P \gM_{MN} + 2 \partial_{(M} \Lambda^P \gM_{N)P} - 2 \partial^P \Lambda_{(M} \gM_{N)P} 
\quad , \quad
\delta_\Lambda (e^{-2d} ) = \partial_P ( \Lambda^P e^{-2d}) \,.
\label{gendiffeo}
\ee
The algebra of generalised diffeomorphisms is closed, and the action invariant, if we impose the section condition $\eta^{MN} \partial_M \otimes \partial_N = 0$. Before applying this requirement, the fields may depend in principle on any of the doubled coordinates. 

Finally, the constant prefactor $G_{DFT}$ is defined formally by $G_{DFT} = G_N \int d^d \tilde x$, where $G_N$ is the spacetime Newton's constant and we have a (perhaps formal) integration only over the dual coordinates. When one solves the section condition by requiring $\tilde\partial^i=0$, the following parametrisation of the generalised metric
\be
\gM_{MN} = \begin{pmatrix}
g_{ij}  -B_{ik} g^{kl} B_{lj}  & B_{ik} g^{kj}  \\
- g^{ik} B_{kj}  & g^{ij} 
\end{pmatrix} \,,
\label{GM}
\ee
together with that of the generalised dilaton, $e^{-2d} = \sqrt{|g|} e^{-2(\phi-\phi_0)}$ (see \cite{Arvanitakis:2016zes} for a discussion of this slightly unconventional choice), reduces the action \eqref{SDFT} to the conventional NSNS sector action\footnote{Our spacetime metric is Lorentzian with mostly plus signature $(-,+,\dots,+)$. Our Ricci scalar is $R = g^{ij} R^k{}_{ikj}$ with Riemann tensor $R^i{}_{jkl} = 2\partial_{[k} \Gamma_{l]j}{}^i + 2 \Gamma_{[k|m}{}^i \Gamma_{l] j}{}^m$, so that the Einstein-Hilbert Lagrangian is $+R$, whereas for a metric of mostly minus signature it would be $-R$.}
\be
S_{NSNS} = \frac{ e^{2\phi_0} }{ 16 \pi G_N }\int d^Dx \sqrt{|g|} e^{-2\phi} \left(R  -\frac{1}{12} H^2 - 4(\nabla \phi)^2 + 4 \nabla^2 \phi\right) \,.
\ee
The second term in the full action $S = S_{DFT} + S_{DWS}$ is the doubled worldsheet action, which we take following \cite{Blair:2013noa}:
\be
S_{DWS} = T \int d^2\sigma \left( \frac{1}{2} \eta_{MN} \dot{X}^M X^{\prime N} 
- \frac{\lambda}{4} ( \gM - \eta)_{MN} X^{\prime M} X^{\prime N}  
- \frac{\tilde\lambda}{4} ( \gM + \eta)_{MN} X^{\prime M} X^{\prime N} \right) \,.
\label{SDWS}
\ee
The worldsheet coordinates are denoted $(\tau,\sigma)$, and $\dot{X} \equiv \partial_\tau X$, $X^\prime \equiv \partial_\sigma X$. Here $\lambda$ and $\tilde \lambda$ are Lagrange multipliers for the Hamiltonian constraints. The algebra of constraints is closed if $\eta^{MN} \partial_M \gM_{PQ} \partial_N \gM_{RS} = 0$ at any two points on the worldsheet \cite{Blair:2013noa}. Picking $\lambda = \tilde\lambda = 1$ corresponds to conformal gauge, in which case $S_{DWS}$ is exactly the Tseytlin action \cite{Tseytlin:1990nb,Tseytlin:1990va}. In what follows, we will make this choice immediately after deriving the equations of motion. The string tension $T$ takes its usual value, $T = 1/2\pi \alpha^\prime$ (we could alternatively have absorbed this into the definition of the coordinates or the generalised metric, but have chosen not to do so).  The action reduces to the conventional string action on imposing the section condition and integrating out $\tilde X_i^\prime$, which is here related to the canonical momenta $P_i$ of $X^i$ by $\tilde X_i = T^{-1} P_i$, so that in conformal gauge
\be
S_{DWS} \rightarrow S_{WS} = \frac{T}{2} \int d^2\sigma \left( g_{ij} \left [ \dot{X}^i \dot{X}^j - X^{\prime i} X^{\prime j} \right] + 2 B_{ij} \dot{X}^i X^{\prime j} \right) \,.
\label{SWS}
\ee
In doing so one may need to worry about boundary terms and zero modes, as discussed in \cite{Blair:2013noa, Berman:2013eva}. (This does not affect the equation of motion of $\gM_{MN}$, nor that of the worldsheet fields if we make the assumption that $\dot{X}^M$ is periodic.) Finally, let us note that one could also include a Fradkin-Tseytlin coupling to the generalised dilaton $d$, however this will not play a role in what follows. 

The equations of motion that follow from the variation of $X^M$ and the Lagrange multipliers in $S_{DWS}$ are then respectively:
\be
\partial_\sigma ( \eta_{MN} \dot{X}^N - \gM_{MN} X^{\prime N} ) + \frac{1}{2} \partial_M \gM_{PQ} X^{\prime P} X^{\prime Q} =0\,,
\label{TEom}
\ee 
\be
\eta_{MN} X^{\prime M} X^{\prime N} = 0 \quad , \quad
\gM_{MN} X^{\prime M} X^{\prime N} = 0 \,.
\label{HamEom}
\ee
The variation of the full action $S$ with respect to the (inverse) generalised metric gives the equation of motion 
\be
\frac{e^{-2d}}{16 \pi G_{DFT} } \mathcal{R}_{MN} + \frac{T}{4} \int d^2 \sigma ( \gM_{MP} \gM_{NQ} - \eta_{MP} \eta_{NQ}  ) X^{\prime P} X^{\prime Q} \delta^{(2D)} (x - X) = 0 \,,
\label{HEom}
\ee
where we have used the fact that the variation of $\gM^{MN}$ must respect the coset condition $\gM_{MP} \eta^{PQ} \gM_{QN} = \eta_{MN}$. The result of the bulk variation, $\mathcal{R}_{MN}$, constitutes the generalised Ricci tensor of DFT, and is written down in appendix \ref{dfteom}. Finally, the equation of motion of the generalised dilaton is $\mathcal{R} = 0$.

We now study \eqref{HEom} in more detail. Reinstating the variation $\delta \gM^{MN}$, it is straightforward to calculate
\be
\begin{split}
\delta \gM^{MN} \frac{1}{2} \left(\gM_{MP} \gM_{NQ} - \eta_{MP} \eta_{NQ}  \right) X^{\prime P} X^{\prime Q} & = 
\delta g^{ij} \left( g_{ik} g_{jl} X^{\prime k} X^{\prime l} - ( \tilde X_i^\prime - B_{ik} X^{\prime k} )( \tilde X_j^\prime - B_{jl} X^{\prime l} )\right) 
\\ &\qquad  + \delta B_{ij}  g^{ik} ( \tilde X_k^\prime - B_{kl} X^{\prime l} ) X^{\prime j} \,.
\end{split}
\label{eomcalc} 
\ee
For the coordinates $X^i$, one can integrate \eqref{TEom} to obtain $\dot{X}^i = \gM^{i}{}_N X^{\prime N} = g^{ij} ( \tilde X_j^\prime - B_{jk} X^{\prime k})$. Using this, one has
\be
\begin{split}
\delta \gM^{MN} \frac{1}{2} \left( \gM_{MP} \gM_{NQ} -\eta_{MP} \eta_{NQ}   \right) X^{\prime P} X^{\prime Q} & = 
\delta g^{ij}  g_{ik} g_{jl} ( -\dot{X}^{k} \dot{X}^l + X^{\prime k} X^{\prime l} )
\\ & \qquad + \delta B_{ij} \dot{X}^i X^{\prime j} \,.
\end{split}
\label{EomContrib}
\ee
It is then easy to see that the equations of motion from varying the ordinary string action \eqref{SWS} with respect to $g_{ij}$ and $B_{ij}$ exactly agree with \eqref{EomContrib}, as is entirely expected. 

\section{F1/pp-wave solutions}
\label{solution}

\subsection{The F1/pp-wave solution and double static gauge} 
\label{solF1pp}

The conventional F1 solution in supergravity \cite{Dabholkar:1995nc, Callan:1995hn} can be found by varying the combined action $S_{NSNS} + S_{WS}$ and imposing static gauge: $X^0 = \tau$, $X^1 = \sigma$. Then one finds a solution representing a string lying in the $X^1$ direction:
\be
\begin{split}
ds^2 & = H^{-1} ( - dt^2 + dz^2 ) + d \vec{x}_8^2 \,,\\
B_{tz} & = (H^{-1} - 1) dt \wedge dz \,, \\
 e^{-2(\phi-\phi_0)} & = H \,,
\end{split} 
\label{F1}
\ee
where $H = 1 + \frac{h}{|\vec{x}_8|^6}$, and we have identified $t \equiv x^0 \equiv X^0$, $z \equiv x^1 \equiv X^1$. The constant $h$ will be determined by the source.

For the Tseytlin string, the static gauge must be supplemented by specifying also some of the dual coordinates. Given that we already know the answer is \eqref{F1}, we first write down the generalised metric:
\be
\footnotesize
\gM_{MN} = \begin{pmatrix} 
H-2 & 0 & 0 & 1-H & 0 & 0 \\
0 & 2-H  & 1-H & 0 & 0 & 0 \\
0 & 1-H & -H & 0 & 0 & 0 \\
1-H & 0 & 0 & H & 0 & 0 \\
0 & 0 & 0 & 0 & \delta_{ij} & 0 \\ 
0 & 0 & 0 & 0 & 0 &  \delta^{ij} 
\end{pmatrix} \,,
\label{gmf1}
\ee 
\normalsize
where we have chosen to order the coordinates $x^M = (t,z,\tilde t, \tilde z, x^i, \tilde x_i)$. (The generalised dilaton is constant, we can take it to be zero by choosing the asymptotic value of the spacetime dilaton in this frame to be $\phi_0=0$.)
%\be
%\begin{split}
%\gM_{00} = - \gM_{11} = H-2 
%\quad , \quad
%\gM^{00}=  - \gM^{11} = - H
%\quad , \quad
%\gM_0{}^1 = \gM_1{}^0 = 1 - H \,.
%\end{split} 
%\ee
Then one finds that the doubled static gauge choice
\be
X^0 = \tau \quad , \quad 
X^1 = \sigma \quad , \quad
\tilde X_0 = - \sigma \quad , \quad
\tilde X_1 = \tau \,,
\label{staticf1}
\ee
solves the equations of motion \eqref{TEom} and \eqref{HamEom}. 
In fact, this worldsheet configuration obeys the duality relation, $\dot{X}^M = \gM^{M}{}_N X^{\prime N}$, with the extra term involving a derivative of the generalised metric in the equation of motion \eqref{TEom} cancelling identically. (Note that, as can be seen in the doubled worldsheet action \eqref{SDWS}, we have two Hamiltonian constraints which generate worldsheet diffeomorphisms \cite{Blair:2013noa}, and which we can use to make the static gauge choice for $X^0$ and $X^1$: the configuration for the remaining coordinates is then determined by the equations of motion.) 

Now, the T-dual of the F1 solution is a pp-wave solution.
\be
\begin{split}
ds^2 & = - H^{-1}  dt^2 + H ( d \tilde z  + (H^{-1} - 1) dt )^2  + d \vec{x}_8^2 \,, \\
B& = 0\,, \\
 e^{-2(\phi-\phi_0)} & = 1 \,.
\end{split} 
\label{pp}
\ee
We see that the doubled worldsheet solution also acts as a source for this solution; the choice of static gauge would be 
\be
X^0 = \tau \quad , \quad 
X^1 = \tau \quad , \quad
\tilde X_0 = - \sigma \quad , \quad
\tilde X_1 = \sigma \,,
\ee
with in this duality frame $\tilde z \equiv x^1 \equiv X^1$ and $z \equiv \tilde x_1 \equiv \tilde X_1$. Intuitively, this is a choice of static gauge for which the string is \emph{oriented in a dual direction}. This allows the doubled worldsheet action to source a particle-like wave solution. 

\subsection{Smearing in the dual directions}

One way to view the doubled solution which reduces to the F1/pp-wave is a wave smeared in the dual directions \cite{Berkeley:2014nza}. Let us see explicitly here how this smearing works, by first solving the equations of motion and then applying the section condition, rather than the other way around. We make use of the generalised metric \eqref{gmf1}, but allow the more general ansatz that the function $H$ can depend on the transverse coordinates and their duals, $H = H(x,\tilde x)$. 
The worldsheet equations of motion are still solved by the configuration \eqref{staticf1}. 

Now we consider the equation of motion of the generalised metric given by \eqref{HEom}. Using the generalised metric \eqref{gmf1} and the expression for $\mathcal{R}_{MN}$ in appendix \eqref{dfteom}, one finds that
\be
\mathcal{R}_{MN} = - \frac{1}{8} \left( \delta_M{}^P \delta_N{}^Q - \gM_{M}{}^P \gM_N{}^Q \right) \Box \gM_{PQ} \,,
\ee
where $\Box \equiv \delta^{ij} \partial_i \partial_j + \delta_{ij} \tilde\partial^i \tilde\partial^j$. 
Then we find that the function $H$ obeys 
\be
\Box H + 16\pi G_{DFT} T \delta( \vec{x}_8 ) \delta ( \vec{\tilde x}_8 ) \delta( \tilde t + z ) \delta ( t - \tilde z ) = 0 \,,
\ee
which is solved by
\be
H = 1 + \frac{16 \cdot 180 G_{DFT} T }{\pi^7 r^{14} } \delta ( \tilde t+z) \delta ( t-\tilde z) \qquad r \equiv \sqrt{ \vec{x}_8^2 + \vec{\tilde{x}}_8^2 } 
\ee
This is localised both in the physical transverse coordinates and their duals. 
(Here we are using the vector notation both where appropriate and inappropriate, so that $\vec{\tilde x}$ is really the covector with indices $\tilde x_i$ and $\vec{\tilde{x}}^2 \equiv \delta^{ij}\tilde x_i \tilde x_j$.)
Hence we next smear this over the dual directions, by arraying centres along $\tilde x_i$ at intervals of $2\pi \tilde R_i$. The harmonic function becomes
\be
H = 1 + \frac{8T}{\pi^3} \frac{ G_{DFT} }{ (2\pi)^8 \tilde R_1 \dots \tilde R_8 } \frac{1}{r^6} \delta(\tilde t + z) \delta ( t - \tilde z)  \quad , \quad r \equiv \sqrt{\vec{x}_8^2} \,.
\label{Harmonic} 
\ee
With our definition of $G_{DFT} = G_N \Pi_i (2\pi \tilde R_i)$ we recover exactly the usual string solution. We could also view the endpoint of this process as taking the limit $\tilde R_i \rightarrow 0$, keeping $G_N$ fixed, so that we shrink away the dual directions completely.

The additional delta functions $\delta(\tilde t + z) \delta ( t - \tilde z)$ appearing in \eqref{Harmonic} are unusual from the point of view of the fundamental string solution, however for the pp-wave solution the $\delta ( t - \tilde z )$ is a consequence of the fact that the wave travels in the $\tilde z$ direction at the speed of light (such a delta function appears explicitly in the Aichelburg-Sexl solution which is sourced by a massless particle action \cite{Aichelburg:1970dh, Ortin:2004ms}). The other suggests that there is a similar interpretation in the frame with physical coordinates $(\tilde t, z)$. Regarding the section condition as applied to delta functions, the latter are of course more properly distributions. Inside an integral, for $f$ some ordinary function, we have $\partial_M \delta (x) f(x) = - \delta(x) \partial_M f(x)$, suggesting that we should always transfer derivatives off the delta function and onto any nearby functions, which will be required to obey the section condition. Then for instance $\eta^{MN} \partial_M \delta \partial_N f = - \eta^{MN} \delta \partial_M \partial_N f = 0$.

\subsection{The vibrating string solution} 
\label{vib}

Having seen that the fundamental string and pp-wave solution correspond to doubled string sources of the DFT eom, let us now extend the discussion to consider possible superpositions. Can we have a solution carrying both momentum and winding in the same direction? As shown in \cite{Dabholkar:1995nc}, the naive superposition of the string and wave solutions is not a true solution of string theory: it cannot be sourced by a fundamental string. However, there is such a background (also constructed in \cite{Callan:1995hn}) which is sourced by a macroscopic fundamental string carrying solely left-moving excitations. The solution in asymptotically flat coordinates is given by
\be
\begin{split}
ds^2 & = - H^{-1} \left( du dv - ( 1-H) \dot{F}^2 dv^2 + 2 ( 1 - H) \delta_{ij} \dot{F}^i dx^j dv \right) 
+ \delta_{ij} dx^i dx^j\,,\\
B & = \frac{1}{2} ( H^{-1} - 1) du\wedge dv + \dot{F}_i ( H^{-1} - 1 ) dv \wedge dx^i \,,\\
e^{-2(\phi-\phi_0)} & = H \,,\\
H(\vec{x}, v)  & =  1 + \frac{Q}{|\vec{x}-\vec{F}(v) |^6} \,,
\end{split} 
\label{vibstr}
\ee
where the vector $\vec{F} = \vec{F}(v)$ gives the profile of the vibrating string in the transverse directions: it is a function solely of the lightcone coordinate $v = t + z$, with the other being $u=t-z$. The generalised metric can be written down in coordinates $X^M = ( u,v,\tilde u, \tilde v, x^i, \tilde x_i)$:
\be
\footnotesize
\gM_{MN} = \begin{pmatrix}
0 & \frac{1}{2} (H-2) & H-1 & 0 & 0 & 0 \\
\frac{1}{2} (H-2) & (H-1) \dot{F}^2 & 0 & (1-H) & \dot{F}_j (1-H) & \dot{F}^j (1-H) \\
H-1 & 0 & - 4 (H-1) \dot{F}^2 & - 2H & 2 \dot{F}_j ( 1 -H) & 2 \dot{F}^j (1-H) \\
0 & 1-H & - 2H & 0 & 0 & 0 \\
0 & \dot{F}_i (1-H) & 2 \dot{F}_i(1-H) & 0 & \delta_{ij} & 0 \\
0 & \dot{F}^i(1-H) & 2 \dot{F}^i(1-H) & 0 & 0 & \delta^{ij}
\end{pmatrix} 
\label{Hvib} 
\ee
\normalsize
The generalised dilaton is constant. We will use the flat metric $\delta_{ij}$ to raise and lower the transverse indices.

Having embedded this solution into DFT, we can proceed similarly to before. Let us examine solutions of the equations of motion with \eqref{Hvib} as our ansatz, assuming that $H= H(x,\tilde x, v)$. We follow \cite{Dabholkar:1995nc} and find the following worldsheet coordinate choice:
\be
\begin{split} 
U &  = ( Rn + a ) \sigma^- + \int^{Rn\sigma^+}\!\!\!\!\!\!\!\!\! dv \dot{F}^2(v) \\
V & = Rn\sigma^+ \\
X^i & = F^i(Rn\sigma^+) 
\end{split}\quad
\begin{split} 
\tilde U & = - \frac{1}{2}Rn \sigma^+ \\
\tilde V & = \frac{1}{2} (Rn+a) \sigma^- - \frac{1}{2} \int^{Rn\sigma^+} \!\!\!\!\!\!\!\!\!dv \dot{F}^2(v) \\
\tilde X_i & = F_i(Rn\sigma^+) \,,
\end{split} 
\ee 
where $a \equiv \int^{2\pi Rn} \dot{F}^2$ is the zero mode of $\dot{F}^2$, and $\sigma^\pm = \tau\pm\sigma$. 
The string winds $n$ times around the direction $z$ which is of radius $R$, and carries momentum $p_z \sim a$ in this direction. (Note that for the usual F1 solution, we implicitly had $n=1$ and $R=1$.)
These solve the equations of motion and Virasoro constraints of the doubled string, both for this background and for flat space, so that one has just $\dot{X}^M = \gM^M{}_N X^{\prime N}$. One sees that both the physical coordinates and the duals have the same vibration profile. 

By picking for instance the $^{uu}$ component of the equations of motion, one finds that\footnote{Note that the form of the generalised metric is quite special in that 
the components $\gM_{iM}$ and $\gM{}^i{}_M$ are the same up to the raising and lowering with the flat transverse metric. 
When we extend the supergravity solution to include dependence on the dual coordinates, we have $\partial_v H = - \dot{F}^i \partial_i H - \dot{F}_i \tilde\partial^i H$. As the structure in the transverse and dual transverse directions is, however, the same, this amounts to merely changing the number of transverse coordinates in the same manner as if one had changed the dimension of spacetime. As the supergravity solution solves the NSNS sector equations of motion for general $D$ \cite{Dabholkar:1995nc}, this guarantees that we can rely on the cancellations that occur in that calculation to reduce the equations of motion to the $\Box H$ term.}
\be
\Box H + 16 \pi G_{DFT} nT \, \delta ( \vec{x} - \vec{F} (v) ) \delta ( \vec{\tilde{x}} - \vec{F}(v) ) \delta ( u - \frac{1}{2} \tilde v ) \delta ( v+ \frac{1}{2} \tilde u ) = 0 \,,
\ee
which is solved by
\be
H = 1 + \frac{16 \cdot 180\, G_{DFT} n T }{\pi^7 r^{14} } \delta ( u - \frac{1}{2} \tilde v ) \delta ( v+ \frac{1}{2} \tilde u ) \qquad r \equiv \sqrt{ ( \vec{x} - \vec{F})^2 + (\vec{\tilde{x}} - \vec{F} )^2 } 
\ee
Smearing over the dual directions then reproduces the standard solution \eqref{vibstr}, and determines the constant $Q$. 

The section condition mandates us to smear over half of the coordinates. The half of the coordinates which one can depend on need not be the ones appearing in the choice of spacetime frame: that is to say, configurations $g_{ij} = g_{ij}( \tilde x)$ are valid in DFT: they obey the section condition but are not even locally geometric in spacetime. For the case of the vibrating string, one might imagine therefore constructing a solution which vibrates solely in the dual directions (or in some combination of dual and physical directions).\footnote{However, it would seem that the ``physical'' directions may be required to have zero radius.} From the DFT point of view, this is related to \eqref{vibstr} by the formal analogue of Buscher duality in the $\vec{x}$ directions. It would be interesting to pursue further this speculative idea. 

% added 
Let us note here that we could similarly consider the F1/pp-wave solution of section \ref{solF1pp} with dependence on dual transverse coordinates. This would then represent a fundamental string (or pp-wave) which is localised in the dual space and smeared in the physical ones. The physical interpretation of this is less clear. Solutions localised in dual directions have been discussed before in DFT in the context of the NS5 brane and KK monopole, where they can be very interestingly connected to worldsheet instanton effects \cite{Jensen:2011jna,Berman:2014jsa}.

The vibrating string solution here can be used as the starting point to construct the multiwound string configuration of Lunin and Mathur \cite{Lunin:2001fv}, which is related by dualities to the D1-D5 system, and also to a supertube configuration involving the exotic $5_2^2$ brane \cite{deBoer:2012ma}. We make some further comments about this in appendix \ref{dstube}.

A couple of comments to conclude this section. In \cite{Dabholkar:1995nc}, care is taken to assume that the radius $R$ of the circle the string wraps is large (compared to the string scale) to avoid ambiguities resulting from quantum effects on the worldsheet. In the doubled theory, a large circle will always be accompanied by the T-dual small circle, and it seems we could equally well work in the duality frame of the latter. This seems a little funny. Notions of scale in the doubled theory are not quite as clear-cut as one might like: for instance, there is no notion of scalar curvature in DFT, the generalised Ricci scalar $\mathcal{R}$ vanishing by the dilaton equation of motion, and there do not appear to be other satisfactory possibilities \cite{Hohm:2011si}. 

We were interested in this solution because it represented the correct form of a solution carrying momentum and winding in the same direction, the naive superposition not being an authentic string theory background. Our approach remained more or less to begin with the supergravity solution and work out what the generalised metric for it must be. It would be interesting to develop a better understanding of how superpositions of basic 1/2-BPS solutions work in DFT: addressing for instance the issue that the generalised metric is non-linear in the spacetime fields, possibly precluding an easy generalisation of the usual harmonic superposition rules. This and a more systematic analysis of DFT backgrounds beyond the 1/2-BPS sector of \cite{Berkeley:2014nza, Berman:2014jsa, Berman:2014hna, Bakhmatov:2016kfn} is left for future work. 

\section{Timelike T-duality and the negative F1}
\label{Tsolution}

\subsection{The negative F1}

We now return to the original F1 solution specified by the generalised metric \eqref{gmf1} and static gauge configuration \eqref{staticf1}.
In the doubled formalism, we can apply a further Buscher duality in the timelike direction. This corresponds to making the following choice of static gauge:
\be
X^0 = -\sigma \quad , \quad 
X^1 = \tau \quad , \quad
\tilde X_0 = \tau \quad , \quad
\tilde X_1 = \sigma \,.
\label{StaticMad}
\ee 
We take $\tilde t \equiv x^0 \equiv X^0$, $t \equiv \tilde x_0 \equiv \tilde X_0$. The string would then seem to be oriented entirely in the dual space. The resulting spacetime is singular:
\be
\begin{split}
ds^2 & = \frac{1}{2-H} ( - d\tilde t^2 + d\tilde z^2 ) + d \vec{x}_8^2 \, \\
B & = \frac{1-H}{H-2} d\tilde t \wedge d\tilde z \, \\
 e^{-2(\phi-\phi_0)} & = | H-2| \,,
\end{split} 
\label{F1t}
\ee
Let us define $\tilde H = 2 - H = 1 - \frac{h}{|\vec{x}_8|^6}$. Then the above configuration can be written as 
\be
\begin{split}
ds^2 & = \tilde H^{-1} (  -d\tilde t^2 + d\tilde z^2 ) + d \vec{x}_8^2 \, \\
B & = ( \tilde H^{-1} - 1) d\tilde t \wedge d\tilde z \, \\
 e^{-2(\phi-\phi_0)} & = | \tilde H | \,.
\end{split} 
\label{negF1}
\ee
This reveals that the background has the same form as the usual F1 solution \eqref{F1}, but with the sign of $\frac{h}{|\vec{x}_8|^6}$ term in the harmonic function flipped. This is the solution one would obtain for a ``negative F1'', i.e. one with negative tension. Negative branes have been recently studied extensively in \cite{Dijkgraaf:2016lym}. Such branes have harmonic functions $\tilde H = 1 - \frac{h}{r^n}$, with a naked singularity at $\tilde H = 0$. This singularity sets the location of a ``bubble'' surrounding the negative brane. One can probe the interior of the bubble beyond the singularity using other (mutually BPS) branes. Within this bubble, physics is supposed to be described by an exotic string theory, with the spacetime signature in the worldvolume directions flipped. In the case of the negative F1, this theory has the usual spacetime signature (as we flip one time and one space direction), but has D-branes with Euclidean worldvolume theories only. One can then regard the negative F1 as the standard F1 of the exotic theory. 

Looking at the solution \eqref{negF1}, we see that for $\tilde H < 0$ (which is inside the bubble), the signature of the worldvolume indeed flips, so that $\tilde t$ becomes spacelike and $\tilde z$ timelike. This seems to also follow naturally from the static gauge configuration \eqref{StaticMad}, where the timelike worldsheet coordinate, $\tau$, is identified with $X^1 \equiv \tilde z$. In other words, the doubled worldsheet configuration knows about the negative string inside the bubble.

Our point of view here would seem to be that the correct description remains the DFT action plus the doubled worldsheet model. 
Ordinarily in double field theory, one way of evading the singularities that appear in \eqref{F1t} is to use a different parametrisation of the generalised metric, in terms of a metric and an antisymmetric bivector field, $\beta^{ij}$, similar to the discussion in generalised geometry in \cite{Grana:2008yw}. One can usually use the generalised Lorentz symmetry to gauge away the bivector in favour of the $B$-field, however in non-geometric settings there may be obstructions to doing so. 

The form of the bivector for the above solution is $\beta^{tz} = H^{-1} - 1$, which can be compared to the original $B$-field, $B_{tz} = H^{-1} - 1$. The negative F1 can be said to couple electrically to $\beta^{ij}$. This is interesting, because it leads to a connection with another type of exotic brane: the solution which couples magnetically to the bivector is the non-geometric $5_2^2$ (see \cite{deBoer:2012ma} for an extensive treatment of this object and exotic branes in general).  

Let us briefly explain what we mean. Two T-dualities transverse to the NS5 brane, which couples magnetically to the $B$-field, lead to the $5_2^2$ brane. This brane is non-geometric in that it is only globally defined up to a T-duality transformation. The solution can be expressed in terms of the bivector field $\beta^{ij}$,  and carries a ``Q-flux'', $Q_i{}^{jk} \sim \partial_i \beta^{jk}$ which is dual to the standard $H$-flux sourced by the NS5 brane. We regard the $5_2^2$ as coupling magnetically to this bivector. (We can think of $\beta^{ij}$ as a zero-form in spacetime, carrying two vector indices corresponding to the special isometry directions of the $5_2^2$.) 

Then, given that the electric dual to the NS5 is the fundamental string, it is natural to wonder what is the corresponding dual of the $5_2^2$: the correct answer is seen to be the configuration given by dualising the F1 on both worldvolume directions \cite{Bergshoeff:2011se, Sakatani:2014hba}. We now recognise this, following \cite{Dijkgraaf:2016lym}, as a negative F1. 

We therefore potentially have two different outlooks on this sort of exotica. From the point of view of \cite{Dijkgraaf:2016lym}, negative branes should be described in conjunction with exotic versions of string and M-theory. Within such a theory, the negative brane is rendered unexotic: it is a standard object. Alternatively, these branes can be thought of as the objects which couple to unusual (mixed-symmetry) tensor fields (as in the analysis of \cite{Bergshoeff:2011se}, such as the bivector, which can be incorporated into (the perhaps also somewhat ``exotic'') duality invariant approaches as reparametrisations of the generalised metric (in some cases, this has only been carried out at the linear level \cite{Bergshoeff:2016ncb}). The resulting spacetime reductions of the more general DFT description will give essentially what is known as ``$\beta$-supergravity'' \cite{Andriot:2011uh, Andriot:2012an, Andriot:2013xca}.

The negative F1 solution has appeared in \cite{Sakatani:2014hba , Blair:2015eba , Park:2015bza}, where it was interpreted as the electric dual to the exotic $5_2^2$ brane in the above manner, and also observed to have negative ADM mass. 
We now learn from \cite{Dijkgraaf:2016lym} that this is expected for a negative brane viewed as an exotic object in a normal string theory. Inside the bubble though, one should view the brane as a conventional, positive mass object in an exotic theory. 

We will now turn our attention to these exotic string theories. In the next subsection, we will review some pertinent results of \cite{Hull:1998vg, Hull:1998ym, Dijkgraaf:2016lym}, before showing how to construct a variant of DFT to describe a subsector of such theories in section \ref{EF1}.

\subsection{Exotic string theories and timelike dualities} 

The exotic string theories which provide the correct physical description inside the bubble can be found by taking timelike T-dualities of the usual type II theories \cite{Hull:1998vg ,Hull:1998ym}. An important point here is that not only do these theories have different spacetime signatures, but the branes they contain can have worldvolume theories of various signatures.
This last fact is vital for realising that to understand them in the duality manifest setting, we will need additionally to study the doubled worldsheet reformulation of a Euclidean string. 

Let us summarise the essential features of these exotic dualities, in order to set the context. In the notation of \cite{Dijkgraaf:2016lym}, we denote each theory as IIA/B$^{\alpha \beta}$ with $\alpha = \pm$ depending on whether the fundamental string is Lorentzian ($+$) or Euclidean ($-$), and $\beta=\pm$ depending on whether D2/D1 branes are Lorentzian ($+$) or Euclidean ($-$). (Note that the terminology Lorentzian/Euclidean is also used in \cite{Dijkgraaf:2016lym} to mean having an odd/even number of timelike directions.) The conventional type II theories are IIA$^{++}$ and IIB$^{++}$. These are related to each other by spatial T-duality. Timelike T-duality maps IIA$^{++}$ to a novel IIB$^{+-}$ theory containing Euclidean D-branes: this is related by spatial T-duality to IIA$^{+-}$ which is in turn timelike T-dual to the ordinary IIB$^{++}$. (In the terminology of Hull's original paper \cite{Hull:1998ym}, IIA$^{+-} \equiv$ IIA$^*$, IIB$^{+-} \equiv$ IIB$^{*}$.)

No change of signature occurs here. However, one can take the strong coupling limit of IIB$^{+-}$ to obtain a different IIB theory, denoted IIB$^{-+}$ (IIB$^\prime$ in \cite{Hull:1998ym}). This contains Euclidean fundamental strings. A timelike T-duality of this theory gives IIA$^{-+}$ on a spatial circle, with the IIA$^{-+}$ having wholly Euclidean spacetime signature (this is Hull's IIA$_E$). In addition, IIB$^{-+}$ on a spatial circle is dual to a IIA$^{-+}$ on a timelike circle: this IIA$^{-+}$ has however two timelike directions in spacetime. Further dualities can then be taken, leading also to IIA$^{--}$ and IIB$^{--}$ theories. Additionally, one can consider uplifts to variants of M-theory, for instance, the Euclidean IIA$^{-+}$ is given by reducing the usual M-theory on a timelike circle, while the IIA$^{+-}$ theory is the reduction on a timelike circle of an M$^-$ theory with Euclidean M2 branes and two timelike directions. 

We have seen above that the doubled formalism, which automatically contains timelike dualities as a possibility, leads directly to the negative brane solution \eqref{negF1}. The theory inside the bubble in this case is IIA/B$^{+-}$ \cite{Dijkgraaf:2016lym}. Indeed, the DFT action describing both the NSNS and RR sectors was shown in \cite{Hohm:2011dv} to contain the IIA$^{+-}$ and IIB$^{+-}$ theories, by choosing different duality frames related by timelike duality. 

We will provide in the next section the description of the NSNS sector of the IIA/B$^{-\pm}$ theories. In this case, T-duality is believed to change the signature of spacetime. 

However, it has been argued in \cite{Malek:2013sp}, at least in the context of $\mathrm{SL}(3)\times \mathrm{SL}(2)$ and $\mathrm{SL}(5)$ U-dualities, that if one works with the generalised metric as the fundamental field (rather than the spacetime fields that it encodes), then timelike duality transformations should not change the signature of spacetime. Instead, after applying the duality transformation one has to use alternative parametrisations of the generalised metric involving antisymmetric vector fields. It seems likely that similar results should hold for the conventional DFT, where the bivector would then be present. Then one can still accommodate different signature theories as distinct subsectors of double or exceptional field theory, once one has fixed the signature of the generalised metric, as for instance in \cite{Blair:2013gqa}. 

Below, we will evade these issues by noticing that, at least in the context of T-duality, signature change only occurs in the presence of Euclidean strings. We will find that the generalised metric is slightly modified in this case. 

\section{Doubled formalism for Euclidean strings}
\label{EF1}

\subsection{Doubled actions for Euclidean strings} 
\label{Eaction}

In this section, we will provide the basic DFT description which applies when strings have Euclidean worldsheets. The Euclidean worldsheet action, following \cite{Dijkgraaf:2016lym} but with opposite sign for the $B$-field, is
\be
S = \frac{T}{2} \int d^2\sigma \left( 
| \gamma |^{1/2} \gamma^{\alpha \beta} \partial_\alpha X^i \partial_\beta X^j g_{ij} 
+ \epsilon^{\alpha \beta} B_{ij} \partial_\alpha X^i \partial_\beta X^j \right) \,.
\label{SEuc}
\ee
We will continue to refer to the worldsheet coordinates as $\sigma^1 \equiv \tau$ and $\sigma^2 \equiv \sigma$. We take $\epsilon^{12} =1$. 
% added
We stress that the worldsheet is Euclidean but the spacetime metric $g_{ij}$ here still has the standard Lorentzian (mostly plus) signature.

To construct a doubled worldsheet action, we follow the steps of \cite{Blair:2013noa}. First, one writes the Lagrangian in the Hamiltonian form $\mathcal{L} = \dot{X} \cdot P - Ham(X,P)$, where $Ham(X,P)$ is the Hamiltonian. Of course, there is an ambiguity in choosing which direction is worldsheet ``time'' now. In doing this, a convenient parametrisation of the Euclidean worldsheet metric is
\be
\gamma_{\alpha \beta} = \Omega \begin{pmatrix} u^2 + \tilde u^2 & \tilde u \\ \tilde u & 1 \end{pmatrix} \,.
\ee
The functions $u$ and $\tilde u$ then enter the action as Lagrange multipliers for the Hamiltonian constraints, and conformal gauge corresponds to $u=1$, $\tilde u=0$. The $O(D,D)$ structure $\eta_{MN}$ and the generalised metric appear naturally in the Hamiltonian: the former is unchanged here, but the generalised metric is found to be
\be
\gM_{MN} = \begin{pmatrix}
-g_{ij}  -B_{ik} g^{kl} B_{lj}  & B_{ik} g^{kj}  \\
- g^{ik} B_{kj}  & g^{ij} 
\end{pmatrix} \,,
\label{EGM}
\ee
with the spacetime metric in the upper left component appearing now with a minus sign. We will comment further on this below.

Finally, one identifies the dual coordinates as $\tilde X_\mu^\prime = T^{-1} P_\mu$, leading to the following Euclidean doubled worldsheet action:
\be
S_{EDWS} = T \int d^2\sigma \left(
\frac{1}{2} \dot{X}^M \eta_{MN} X^{\prime N} 
- \frac{u}{2} \gM_{MN} X^{\prime M} X^{\prime N} 
- \frac{\tilde u}{2} \eta_{MN} X^{\prime M} X^{\prime N} 
\right)\,.
\label{EDWS}
\ee
The extra minus sign in the generalised metric \eqref{EGM} has interesting consequences. If we raise the indices on the generalised metric with $\eta_{MN}$, the result is no longer the inverse of the generalised metric, but minus it. That is, if we define $\gM^{MN} \equiv \eta^{MP} \eta^{NQ} \gM_{PQ}$ then $\gM^{MN} \gM_{NP} = - \delta^M_P$. 

This seems somewhat unusual. 
The generalised metric is now not an element of $O(D,D)$. However, $\pm i \gM$ is an element of the complexified duality group $O(D,D;\mathbb{C})$. We see also that as $\gM \eta^{-1}$ squares to minus one, rather than plus one, it provides a complex structure rather than a product structure.

Buscher dualities now have the effect of seemingly changing the spacetime signature. This is easy to see if there is no B-field. Then under a Buscher duality
\be
\gM_{MN} = \begin{pmatrix} - g & 0 \\ 0 & g^{-1} \end{pmatrix} \rightarrow \begin{pmatrix} g^{-1} & 0 \\ 0 & - g \end{pmatrix}
\equiv \begin{pmatrix} - \tilde g & 0 \\ 0 & \tilde g^{-1} \end{pmatrix} \,,
\ee 
and the dual metric is $\tilde g = - g^{-1}$, with the appearance of the new minus sign interpreted as having the effect of flipping the metric signature. (One can check that the Buscher rules implied by the transformation of \eqref{EGM} agree with \cite{Hull:1998ym,Dijkgraaf:2016lym}.) Note that the signature of the generalised metric is $(D,D)$, rather than $(2,2(D-1))$. We shall discuss this in more detail in the subsequent subsection. 

The next step to take is to produce the corresponding double field theory action. Such an action will have equations of motion which correspond to the beta functional equations of the worldsheet action \eqref{EDWS}. One way to obtain the action, without actually calculating the latter, would be to use an analytical continuation of the Wick-rotated Lorentzian action to obtain the Euclidean action, as used to find the spacetime supergravity actions in \cite{Dijkgraaf:2016lym}. This would seem to involve $g \rightarrow - ig$, $B \rightarrow B$, and $\gM_{MN} \rightarrow - i \gM_{MN}$.

However, it is also very simple to use DFT methods directly. The conventional DFT action is entirely fixed by requiring invariance under the local symmetry of the theory. This means searching for an object quadratic in derivatives of the generalised metric and generalised dilaton, which transforms as a scalar under generalised diffeomorphisms, which are given still by \eqref{gendiffeo} (modulo the section condition, also unchanged). The only change here relative to the calculation presented in \cite{Hohm:2010pp} is the fact that $\gM^{MN}$ is now minus the inverse of the generalised metric \eqref{EGM}. This additional minus sign has the effect of changing the sign of two of the coefficients in \eqref{RDFT}. 

The resulting doubled theory that we obtain could be called DFT$^-$, using the language of \cite{Dijkgraaf:2016lym}. The minus indicates that (doubled) fundamental strings have Euclidean worldsheets. The conventional DFT would then be DFT$^+$. 

The generalised Ricci scalar giving the Lagrangian of DFT$^-$ is found to be
\be
\begin{split}
\mathcal{R}^- & = 4 \gM^{MN}\partial_{M}\partial_{N}d
-\partial_{M}\partial_{N}\gM^{MN} 
-4\mathcal{H}^{MN}\partial_{M}d\,\partial_{N}d
+ 4 \partial_M \gM^{MN}  \,\partial_Nd\\
&-\frac{1}{8}\gM^{MN}\partial_{M}\gM^{KL}\,
\partial_{N}\gM_{KL}+\frac{1}{2}\gM^{MN}\partial_{M}\gM^{KL}
\partial_{K}\gM_{NL}\,.
\end{split}
\label{REDFT}
\ee
On solving the section condition as $\tilde \partial^i =0$ and inserting the parametrisation \eqref{EGM}, this reduces to the spacetime Lagrangian
\be
\mathcal{R}^- \stackrel{\tilde \partial^i=0}{\longrightarrow} 
R + \frac{1}{12} H^2 - 4 ( \nabla\phi)^2 + 4  \nabla^2 \phi \,.
\ee
Here $R$ is the spacetime Ricci scalar and $H$ is the field strength of the $B$-field.\footnote{If the metric $g$ instead has mostly minus signature, as can happen by acting with dualities, then the appropriate Einstein-Hilbert term is $-R$ and $\mathcal{R}^-$ in fact reduces to \emph{minus} the expected (exotic) supergravity Lagrangian, consistent with the relationship between the actions for ``spacetime mirror'' theories obtained by flipping the signature in \cite{Dijkgraaf:2016lym}.} The sign of the $H^2$ term is minus what it is in the usual NSNS action. This is expected for the exotic supergravities of \cite{Hull:1998ym, Dijkgraaf:2016lym}, with the sign of the $B$-field kinetic term non-standard in the IIA/B$^{-\pm}$ theories. Here we have shown that this can be naturally accommodated in double field theory when one takes into account that the generalised metric obeys $\gM^{MN} \gM_{NP} = - \delta^M_P$ leading to the alternative parametrisation \eqref{EGM}. The construction of the worldsheet action \eqref{EDWS} was a convenient way to discover this fact.

\subsection{Concerning the change of signature}
\label{Esig} 

\subsubsection{Generalised vielbeins} 

Let us focus on the subtleties regarding achieving signature change via Buscher dualities in double field theory. 
First, let us introduce the decomposition of the generalised metric in terms of a generalised vielbein:
\be
\gM_{MN} = E_M{}^\alpha \hat{\gM}_{\alpha \beta} E^\beta{}_N \,,
\label{gv} 
\ee
where $\alpha$ is a doubled flat index, and the flat generalised metric is
\be
\hat{\gM}_{\alpha \beta} = \begin{pmatrix} - h_{t,s} & 0 \\ 0 & h_{t,s}^{-1} \end{pmatrix} 
\quad , \quad 
h_{t,s} = \mathrm{diag}\,( \underbrace{-1,\dots,-1}_t , \underbrace{+1,\dots,+1}_s )\,.
\ee
We see that this means the signature of $\gM_{MN}$ is indeed $(D,D)$. 
The parametrisation \eqref{EGM} is obtained by taking the generalised vielbein to be
\be
E^\alpha{}_M = \begin{pmatrix} e & 0 \\ - e^{-T} B & e^{-T} \end{pmatrix} \,,
\ee 
where we are using matrix notation for $e^{a}{}_i$ a vielbein for the spacetime metric of signature $(t,s)$, $g_{ij} = e_i{}^a (h_{t,s})_{ab} e^b{}_j$, i.e. $g = e^T h e$. 

The generalised vielbein transforms under local generalised Lorentz transformations, $E^\alpha{}_M \rightarrow \Lambda^\alpha{}_\beta E^\beta{}_M$, which by definition preserve $\hat{\gM}_{\alpha \beta}$ in addition to the $O(D,D)$ structure with flat indices,
\be
\hat\eta_{\alpha \beta} = \begin{pmatrix} 0 & I \\ I &  0 \end{pmatrix} \,.
\ee
We denote the group of such transformations simply by $H$. We show in appendix \ref{genlor} that $H$ is a somewhat unusual real form of $O(D,\mathbb{C}) \times O(D,\mathbb{C})$.

When we act with a transformation $P \in O(D,D)$ on the generalised vielbein, we will generically need to simultaneously apply a compensating $H$-transformation in order to read off the transformations of the spacetime fields:
\be
\tilde E^\alpha{}_M = \Lambda^\alpha{}_\beta E^\beta{}_N P^N{}_M \,.
\label{compensating}
\ee 

\subsubsection{A simple example}

To illustrate the apparent change of signature, let us focus on the simple case of $D=1$ and dualise on a spacelike circle in DFT$^-$. 
The generalised metric involves just a single metric component,
\be
\gM_{MN} = \begin{pmatrix} - g & 0 \\ 0 & g^{-1} \end{pmatrix} \,.
\ee
We decompose the metric $g = e^2 = e(+1)e$ in terms of an einbein. This metric has positive signature. The signature of the generalised metric is $(1,1)$, so that the decomposition \eqref{gv} is in terms of 
\be
\hat{\gM}_{\alpha \beta} = \begin{pmatrix} -1 & 0 \\ 0 & 1 \end{pmatrix} 
\quad , \quad
E^\alpha{}_M = \begin{pmatrix} e & 0 \\ 0 & e^{-1} \end{pmatrix} \,.
\ee
If we act with a Buscher duality, $P^N{}_M = \footnotesize\begin{pmatrix} 0 & 1 \\ 1 & 0 \end{pmatrix}$, we obtain the new generalised metric
\be
\tilde\gM_{MN} =
\begin{pmatrix} - \tilde g & 0 \\ 0 & \tilde g^{-1} \end{pmatrix} = 
\begin{pmatrix} g^{-1}  & 0 \\ 0 & - g \end{pmatrix} \quad \Rightarrow \tilde g = - g^{-1} \,.
\ee
In order to obtain the new generalised vielbein in the form
\be
\tilde E^\alpha{}_M  = \begin{pmatrix} \tilde e & 0 \\ 0 & \tilde e^{-1} \end{pmatrix} \,,
\ee
we have to apply a compensating transformation as in \eqref{compensating}, which is given by
\be
\Lambda^\alpha{}_\beta = \begin{pmatrix}0 & \tilde e e \\ \tilde e^{-1} e^{-1} & 0 \end{pmatrix} \,,
\ee
which automatically preserves $\eta_{\alpha \beta}$. For it to also preserve $\hat{\gM}_{\alpha \beta}$, we need $\tilde e^2 = - e^{-2}$. Although this is consistent with $\tilde g = - g^{-1}$, it also means that $\tilde e = (\pm i) e^{-1}$, so both the new vielbein $\tilde e$ and the transformation $\Lambda^\alpha{}_\beta$ are imaginary. 

As we required $\Lambda$ to preserve the form of $\hat{\gM}_{\alpha \beta}$, the transformed generalised metric has the same signature, and in particular so naively does $\tilde g$ in the sense that $\tilde g = \tilde e^2 = \tilde e (+1) \tilde e$. If we work with a real spacetime vielbein $\tilde e \rightarrow \pm i \tilde e$, then we have $\tilde g = - \tilde e^2=\tilde e(-1) \tilde e$ and this has the opposite signature. 

Alternatively, one could view this appearance of imaginary transformations and vielbeins as a genuine obstruction to the duality being performed. Although the generalised metric remains real, we cannot guarantee that there will not be further issues for instance when fermions are included, which couple to the vielbein. 

In this case, one possible reinterpretation would be the following. We instead argue that the duality transformation must be accompanied by changing the signs of the entries of the generalised metric $\hat{\gM}_{\alpha \beta}$: this is equivalent to taking the compensating transformation $\Lambda$ to be real and given by
\be
\Lambda^\alpha{}_\beta = \begin{pmatrix} 0 & 1 \\ 1 & 0 \end{pmatrix} \,.
\ee
This no longer preserves $\hat\gM_{\alpha\beta}$ and instead amounts to a change of basis of the generalised tangent space. Then the generalised metric
\be
\gM_{MN} = \begin{pmatrix} -g&0\\0&g^{-1} \end{pmatrix} 
 = \begin{pmatrix} e & 0 \\0 & e^{-1} \end{pmatrix} 
\begin{pmatrix} -1 & 0 \\ 0 & 1 \end{pmatrix} 
\begin{pmatrix} e & 0 \\0 & e^{-1} \end{pmatrix} 
\ee
with $g = e(+1) e$, is Buscher dual to the generalised metric 
\be
\tilde \gM_{MN} = \begin{pmatrix} -\tilde g&0\\0& \tilde g^{-1} \end{pmatrix} 
 = \begin{pmatrix} \tilde e & 0 \\0 &\tilde e^{-1} \end{pmatrix} 
\begin{pmatrix} 1 & 0 \\ 0 & -1 \end{pmatrix} 
\begin{pmatrix} \tilde e & 0 \\0 & \tilde e^{-1} \end{pmatrix} 
\ee
with opposite signature in spacetime, so $\tilde g = \tilde e ( - 1) \tilde e$. However, we now have the relationship $\tilde e = e^{-1}$ implying again $\tilde g = - g^{-1}$. Technically now this duality is a map between two different theories defined by the positioning of the $+1$ and $-1$ in $\hat{\gM}_{\alpha \beta}$. 

Note that the problems exhibited in the above example cannot be circumvented by introducing a $B$-field or a bivector, as we do not have enough dimensions.  

In the case of simultaneous Buscher dualities on multiple directions, similar issues arise if one attempts to take the transformed vielbein to be proportional to the original one, $\tilde e^a{}_\mu = \lambda^a{}_b e^b{}_\mu$. One finds that for the compensating transformation to preserve $\delta_{\alpha\beta}$ that $\lambda^T h\lambda = - h$, where $h$ is the flat Minkowski metric with arbitrary $(t,s)$ signature: this inevitably means that the norm squareds of the rows of $\lambda$ have to be negative, and so the relationship between the vielbeins will not be real, unless one forcibly changes the signature. Here one might be able to introduce both a $B$-field and bivector to avoid this. However, one will be left with the issue for the situation where one carries out a single Buscher duality. 

We point out the generalised metric of DFT$^-$ always has signature $(D,D)$. It is only the spacetime signature that changes when one simultaneously performs a duality transformation and a change of generalised tangent space basis. 

\subsubsection{Using the complex coset}

We can also study the effects of the duality transformation using the complexified coset. An element of $O(D,D;\mathbb{C}) / O(D;\mathbb{C}) \times O(D;\mathbb{C})$ can be parametrised as in \eqref{GM} in terms of a metric $g$ and $B$-field $B$, both viewed as $D \times D$ matrices but with complex entries. 
If we restrict to purely imaginary elements, then we need to take $g \rightarrow - i g$ with $g$ now real, and $B$ real: such elements are then of the form
\be
i \begin{pmatrix} - g - B g^{-1} B & B g^{-1} \\ - g^{-1} B & g^{-1} \end{pmatrix} \,.
\ee
This shows that we can take $i\gM_{MN}$ to be a purely imaginary element of the complexified coset.

Let us look again at our simple $D=1$ example, with 
\be
i\gM_{MN} = \begin{pmatrix} - i g & 0 \\ 0 & i g^{-1} \end{pmatrix} = 
E_M{}^\alpha \delta_{\alpha \beta} E^\beta{}_N 
\ee 
where the flattened generalised metric is now the identity, $\hat{\gM}_{\alpha\beta} = \delta_{\alpha \beta}$, and the generalised vielbein is
\be
E^\alpha{}_M = \begin{pmatrix} ( - i)^{1/2} e & 0 \\ 0 &  i^{1/2} e^{-1} \end{pmatrix} \,.
\ee
We act again with a Buscher transformation to find that the transformed generalised vielbein and associated compensator transform $\Lambda \in O(1;\mathbb{C}) \times O(1;\mathbb{C})$ is
\be
\tilde E^\alpha{}_M = \begin{pmatrix} \pm i^{1/2} e^{-1} & 0 \\ 0 & \pm (-i)^{1/2} e \end{pmatrix} 
\quad , \quad \Lambda^\alpha{}_\beta = \begin{pmatrix} 0 & \pm 1 \\ \pm 1 & 0 \end{pmatrix} \,.
\ee
The advantage now is that there are no issues regarding viewing $\Lambda$ as an actual element of $H$. 

The main result is the same though: we find that $\tilde e = \pm i e^{-1}$ so that the transformed vielbein is imaginary and we have again $\tilde g = \tilde e^2 = - g^{-1}$. When we restrict to the appropriate real form, we take the vielbein to be real by absorbing the $\pm i$ into a change of signature in spacetime.

The conclusion from this point of view is that the complexified generalised metric, $i\gM_{MN}$ always has positive definite signature. However, after carrying out the Buscher transformation, one is led to restrict to a real form in spacetime which has a different spacetime signature.

\subsection{Euclidean strings are timelike waves} 
\label{Esoln} 

We conclude our discussion of DFT$^-$ by writing down its fundamental string solution and its T-duals. 
The supergravity configuration corresponding to a Euclidean F1 can be found by taking the M2 solution of 11-dimensional supergravity and reducing on a timelike circle. The result is
\be
\begin{split} 
ds^2 & = H^{-1} \left( (dz_1)^2 + (dz_2)^2 \right) + d \vec{x}_8^2 \,,\\
B & = ( H^{-1} - 1 ) dz_1 \wedge dz_2 \,,\\
e^{-2(\phi-\phi_0)} & = H \,.
\end{split}
\label{EF1SUGRA} 
\ee
Here $z_1$ and $z_2$ are the Euclidean worldvolume directions. The harmonic function is $H= 1+\frac{h}{r^6}$. 
The generalised metric (leaving out the transverse directions) is
\be
\footnotesize
\gM_{MN} = \begin{pmatrix} H-2 & 0 & 0 & 1-H \\ 0 & H-2 & H-1 & 0 \\ 0 & H-1 & H & 0 \\ 1-H & 0 & 0 & H \end{pmatrix} \,.
\label{HEF1}
\ee
The generalised dilaton is again constant. 
We can take a Buscher dual on $z_2$, say. Then we get 
\be
\begin{split} 
ds^2 & = - [ - H^{-1} (dz_1)^2 + H \left( d\tilde z_2 + (H^{-1} - 1) dz_1 \right)^2 ]  + d \vec{x}_8^2 \,,\\
B & = 0\\
e^{-2(\phi-\phi_0)} & = 1 \,.
\end{split}
\label{EppSUGRA} 
\ee
The part of the metric in the brackets corresponds to a pp-wave travelling in the $\tilde z$ direction. The overall minus sign implies that this should be considered the timelike direction, which is in accord with the general expectations of \cite{Dijkgraaf:2016lym}. In the sector with Euclidean fundamental strings, a IIA theory on a spacelike circle is dual to IIB on a timelike circle, and vice versa. Thus the T-dual of the Euclidean fundamental string solution is a wave in a timelike direction. 

The solution that is obtained by further dualising on $z_1$ is 
\be
\begin{split} 
ds^2 & = -\tilde H^{-1} ( (d\tilde z_1)^2 + ( d\tilde z_2)^2)  + d \vec{x}_8^2 \,,\\
B & = ( \tilde H^{-1} - 1) d \tilde z_1 \wedge d\tilde z_2\\
e^{-2(\phi-\phi_0)} & = \tilde H \,.
\end{split}
\label{negEF1SUGRA} 
\ee
Here $\tilde H = 2 - H = 1-\frac{h}{r^6}$. This is a negative Euclidean F1. There are now two timelike directions, $\tilde z_1, \tilde z_2$. However, inside the bubble these will both become spacelike.

The configuration \eqref{EF1SUGRA} can be seen to be a solution of the equations of motion of $S = S_{DFT^-} - S_{EDWS}$. 
Note the relative minus sign. This can be absorbed in redefining $S_{EDWS} \rightarrow - S_{EDWS}$, suggesting that this is then the correct Euclidean doubled worldsheet action for a positive tension Euclidean doubled string. The equations of motion in conformal gauge give
\be
\frac{e^{-2d}}{16\pi G_{DFT}} \mathcal{R}_{MN}^-  
- \frac{T}{4} \int d^2 \sigma ( - \gM_{MP} \gM_{NQ} - \eta_{MP} \eta_{NQ}  ) X^{\prime P} X^{\prime Q} \delta^{(2D)} (x - X) =0 \,,
\ee
where the generalised Ricci tensor $\mathcal{R}_{MN}^-$ is defined in appendix \ref{dfteom}. We also have to solve the constraints $\eta_{MN} X^{\prime M} X^{\prime N} = 0 = \gM_{MN} X^{\prime M} X^{\prime N}$, and the equations of motion of the worldsheet coordinates. One can check that with the generalised metric given by \eqref{HEF1}, the configuration
\be
X^1 = \tau \quad , \quad 
X^2 = -\sigma \quad , \quad
\tilde X_1 = \sigma \quad , \quad
\tilde X_2 = \tau \,,
\ee
is a solution, with $\dot{X}^M = \gM^M{}_N X^{\prime N}$.  

If we had used instead the original $S_{EDWS}$, with the opposite sign, we would have obtained directly the solution with $H = 1 - \frac{h}{r^6}$, i.e. that of the negative Euclidean string. This would arise by reducing a negative M2 brane on a timelike direction. We note as well that the Euclidean (undoubled) worldsheet action \eqref{SEuc} is equivalent to the Nambu-Goto action $S_{NG} = + T \int d^2\sigma \sqrt{\mathrm{det}\hat{\gamma}}$, where $\hat{\gamma}$ is the induced Euclidean metric on the worldsheet. The conventional Nambu-Goto action has a minus sign. This further suggests interpreting the action \eqref{SEuc} as actually corresponding to a negative tension Euclidean string.

\section{Conclusions} 
\label{conclusions} 

In this paper, we have used the single framework of the doubled formalism to study all at once familiar fundamental strings, pp-waves and less familiar negative strings. Our original goal was to better understand the origin of the doubled wave solution of \cite{Berkeley:2014nza}, by showing that the equations of motion of double field theory could - and should, in this case - be sourced by a doubled worldsheet action. This worked perfectly, using the Tseytlin action as an example - we show in the appendix that Hull's doubled everything action could also be used. We saw the nice result that the doubled sigma model can source both the F1 and a wave, in different duality frames, depending on the orientation of the string in static gauge, with the pp-wave resulting from the string pointing in a dual direction. 

We took seriously the possibility of carrying out a timelike duality within the doubled formalism. We saw that the configuration dual to the F1 on both worldvolume directions could be regarded as a ``negative string.'' We suggested that doubled actions are a natural setting to study these negative branes. Such branes are surrounded by a bubble containing exotic string theories, as described in \cite{Dijkgraaf:2016lym, Hull:1998vg, Hull:1998ym}. We then focused on the subsector of such theories which contain fundamental strings with Euclidean worldvolumes, and showed how the NSNS sector of these theories can be described in a novel variant of DFT (which we called DFT$^-$) using a modified generalised metric.

This DFT$^-$ could certainly be developed further along the lines of the conventional DFT, for instance by including the RR sector and supersymmetry \cite{Hohm:2011dv, Jeon:2012hp}. One might also wonder about Scherk-Schwarz compactifications of DFT$^-$ and heterotic strings, which can be naturally accommodated in DFT \cite{Hohm:2011ex,Grana:2012rr, Siegel:1993th}. The geometry and relationship to fluxes presumably follows with simple modifications to the usual case \cite{Siegel:1993th, Jeon:2011cn, Hohm:2011si, Berman:2013uda, Geissbuhler:2013uka}. The mathematical structures that appear in the associated generalised geometry \cite{Gualtieri:2003dx} may also be interesting. From a worldsheet point of view, the connection with the ``metastring'' approach \cite{Freidel:2015pka} could be of interest. 

One interesting aspect of this analysis that we want to reiterate is that the negative F1 solution can also be regarded as the electric dual to a different sort of exotic brane, the $5_2^2$. 
The latter couples magnetically to a bivector field, carrying a magnetic ``Q-flux'', while the negative F1 couples electrically to this field. The appearance of bivector fields in the parametrisation of the generalised metric may be closely linked to the appearance of exotic string theories. This suggests the links between DFT, $\beta$-supergravity \cite{Andriot:2011uh, Andriot:2012an, Andriot:2013xca} and such exotic string theories could be investigated further.

Another reason to further study this interplay is provided by the results of \cite{Malek:2013sp} on timelike U-dualities in the exceptional field theory (EFT) setting, which appear to rule out signature changes in favour of introducing (generalisations of) the bivector. In this paper, we suggested that this can be partially circumvented in DFT by using an alternative generalised metric derived from the Euclidean string worldsheet action: this seemed more naturally to live in a complexification of the coset $O(D,D)/O(D) \times O(D)$. A more thorough analysis of the associated subtleties, and how they relate to the introduction of the bivector, may cast further light on these issues. 

Understanding this will be necessary in order to establish the status of these exotic theories in the U-duality framework. 
One starting point could be to return, as in \cite{Berman:2010is}, to the M2 worldvolume approach first attempted in \cite{Duff:1990hn}, in order to find the generalised metric appropriate to the theory with Euclidean M2 branes. Then rather than the usual U-duality invariant exceptional field theory \cite{Berman:2010is,Hohm:2013pua}, this may lead to (perhaps multiple notions of) an EFT$^-$, which cannot be related to the conventional (EFT$^+$) theory by duality. 

Another starting point would be to note that the IIB$^{+-}$ theory is S-dual to IIB$^{-+}$, and so one should in principle be able to describe this by taking the $\mathrm{SL}(2) \times \mathbb{R}^+$ EFT \cite{Berman:2015rcc} and changing the coset to $\mathrm{SL}(2)/\mathrm{SO}(1,1)$. This EFT will also describe an M-theory/IIA section. As one only has access to S-duality in this EFT, this might provide some interesting pointers as how different signature string and M-theories can be accommodated in an EFT, without having introduced more complicated dualities. 

A final suggestion would be to consider timelike reductions from EFT to DFT, along the lines of \cite{Thompson:2011uw}, given that the Euclidean string results from reducing the M2 brane on a timelike direction. 

Perhaps the central trick of this paper was that we drew conclusions about (doubled) spacetime theories by referring back to the (doubled) worldsheet theories. We therefore note that a complication in establishing the possibility of an EFT$^{-}$ starting from M2 worldvolume theory is that there are a number of different signature versions of M-theory available, some of which can accommodate two notions of M2 branes with different signatures, meaning that there may be several distinct notions of EFT$^\pm$.

Indeed, we may notice the observation in \cite{Dijkgraaf:2016lym} that there may be several exotic theories of each type. This could be reflected by there potentially being a number of versions of EFT, determined by the (fixed) signature of the generalised metric. Then, following \cite{Malek:2013sp},  dualities which appear to change the spacetime signature in fact lead to theories with bivector/trivector fields, which effectively take the place of a conjectured exotic theory with a different signature. This exotic theory that has been replaced here may still exist in its own right in a distinct subsector of the space of all theories.
This viewpoint might also make sense from the generalised geometry point of view, where the spacetime manifold should be fixed, and the generalised metric appears on an extended tangent bundle, leading to the reformulations of supergravity developed in \cite{Coimbra:2011nw}. One mostly aesthetic objection to this line of thinking is that what was supposed to be a framework for unifying different theories now in fact apparently turns into an alarming proliferation of them.
We intend to study this situation in future work. 

Let us now give some final thoughts about other directions that could be pursued. The paper \cite{Arvanitakis:2016zes} reformulated black hole thermodynamics in DFT. The example of the black fundamental string solution was studied in some detail, and the configuration obtained from it by duality along both time and space was written down. 
%The solution is
%\be
%\begin{split} 
%ds^2 & = \tilde H^{-1} \left( - d \tilde t^2 + W d \tilde z^2 \right) + W^{-1} dr^2 + r^2 d\Omega_7^2\,, \\
%B & = \alpha^{-1} (1-\tilde H^{-1}) d\tilde t \wedge d\tilde z\,,\\
%e^{-2\phi} & = |\tilde H| \,,
%\end{split} 
%\ee
%where $\tilde H = 1 - r_+^6/r^6$, $W = 1 - ( r_+^6 - r_-^6)/r^6$ and $\alpha = r_+^3/r_-^3$, with $r_+ \rightarrow r_-$ the extremal limit. 
This solution is a black negative string, with the usual horizon now hidden beyond the naked singularity at $\tilde H = 0$. 
%We see again that inside the bubble, $\tilde z$ plays the role of the time coordinate. 
The general properties of non-extremal negative solutions are likely quite exotic as well, and could perhaps be studied from this approach.

The other aspect of the doubled string solution which we addressed in this paper involved the embedding into DFT of the solution corresponding to an oscillating string \cite{Dabholkar:1995nc, Callan:1995hn}. This was of interest as it clarifies how one can construct configurations corresponding to a string carrying both momentum and winding in the same direction. It may be of interest to further explore such solutions, including whether there is a genuine notion of a string vibrating in the dual space (giving some sort of doubled supertube, puffed up in the dual space), and the links to exotic branes \cite{deBoer:2012ma}.

Lastly, an obvious extension of our work would be to attempt to source the equations of motion of DFT or EFT with other duality invariant brane actions, with a view towards for instance the solutions analysed in \cite{Berkeley:2014nza, Berman:2014jsa, Berman:2014hna, Bakhmatov:2016kfn}.
The hard part here is constructing such actions, given that U-duality maps branes of different worldvolume dimension into each other, however this is an interesting problem for the duality manifest approach. Some recent papers addressing this are \cite{ Duff:2015jka, Sakatani:2016sko}.

\section*{Acknowledgements} 

I would like to thank Alex Arvanitakis, David Berman and Daniel Thompson for useful discussions.
I am supported in part by the Belgian Federal Science Policy Office through the Interuniversity Attraction Pole P7/37 ``Fundamental Interactions'', and in part by the ``FWO-Vlaanderen'' through the project G.0207.14N and by the Vrije Universiteit Brussel through the Strategic Research Program ``High-Energy Physics''.

\appendix 

\section{Supplementary results: worldsheet} 
\label{appws} 

\subsection{Hull's doubled string as the source} 
\label{hullsource} 

One could also take the doubled worldsheet action due to Hull \cite{Hull:2004in}, which we write here in the ``doubled everything'' form \cite{Hull:2006va} as 
\be
S_{Hull} = - \frac{T}{4} \int d^2 \sigma \sqrt{|\gamma|} \gamma^{\alpha \beta} \gM_{MN} \partial_\alpha X^M \partial_\beta X^N \,,
\ee 
where $\gamma_{\alpha \beta}$ is the worldsheet metric, supplemented by the constraint 
\be
\partial_\alpha X^M = \gM^M{}_N \epsilon_{\alpha}{}^\beta \partial_\beta X^N\,.
\ee
Note that the normalisation of the action is $T/4$ rather than the usual $T/2$. 
In conformal gauge, the action becomes
\be
S_{Hull} = \frac{T}{4} \int d^2 \sigma \gM_{MN} \left( \dot{X}^M \dot{X}^N - X^{\prime M} X^{\prime N}\right) \,,
\ee
with the constraint 
\be
\dot{X}^M = \gM^M{}_N X^{\prime N} \,,
\label{Hullcon} 
\ee
and the additional equations of motion of the worldsheet metric:
\be
\gM_{MN} \partial_\alpha X^M \partial_\beta X^N - \frac{1}{2} \gamma_{\alpha \beta} \gM_{MN} \gamma^{\gamma \delta} \partial_\gamma X^M \partial_\delta X^N = 0 \,.
\ee
This formulation is equivalent to the Tseytlin string \cite{Berman:2007xn}. 

The equation of motion of the generalised metric following then from the joint action $S = S_{DFT} + S_{Hull}$ in conformal gauge is then
\be
\begin{split}
\frac{e^{-2d}\mathcal{R}_{MN} }{16\pi G_{DFT} } 
 + \frac{T}{8} \int d^2 \sigma \left( \gM_{MP} \gM_{NQ} - \eta_{MP} \eta_{NQ} \right)  \left( \dot{X}^P \dot{X}^Q - X^{\prime P} X^{\prime Q} \right) \delta^{(2D)}(x-X) = 0 \,.
\end{split} 
\ee
Using \eqref{eomcalc} and the constraint \eqref{Hullcon} to write everything in terms of $\dot{X}$ and $X^{\prime}$, one finds that the worldsheet contribution matches exactly that which arises from the conventional sigma model, \eqref{SWS}.

One can also check for instance that the doubled F1/pp-wave configuration solves the above equations and constraints, showing that it can be considered to be sourced by the Hull action.

\subsection{Tseytlin in split form} 
\label{splitTseytlin} 

The following result was obtained in the course of writing this paper, and although we do not actually use it in the main text we record it here as it may be of interest. 
We note that it also follows from the older results of \cite{Schwarz:1993mg}, which considered a Tseytlin-like action for the heterotic string, and the more recent work of \cite{Driezen:2016tnz}. Here we make explicit the relationship to the standard DFT parametrisations. 

For coupling to the full DFT action, it is convenient to work with doubled worldsheet models where all coordinates are doubled. One can also double only some subset, in which case, the appropriate form of the DFT action will be the Kaluza-Klein-inspired one presented in \cite{Hohm:2013nja}. 
One might be interested in this as it is similar to but simpler than the full EFT framework \cite{Hohm:2013pua}, or alternately one might wish to restrict to the situation where we genuinely only double compact directions. In this subsection, we give the appropriate split form of the Tseytlin action that would couple to this version of DFT. 

Take the usual $O(D,D)$ invariant DFT, with generalised metric $\hat\gM_{\hat M \hat N}$ (where we have written the field and indices with hats to make the decomposition clearer). Split the $O(D,D)$ coordinates $X^{\hat M} = ( X^\mu, \tilde X_\mu , X^M)$ where $M$ is an $O(d,d)$ index. We assume everything is independent of $\tilde X_\mu$.  Then parametrise the generalised metric as \cite{Hohm:2013nja}
\begin{align}
\label{KK_parametrisation} 
\hat \gM_{\mu \nu} & = g_{\mu \nu} + g^{ \rho \sigma} C_{\mu  \rho} C_{\nu \sigma} + \gM_{M N} A_\mu{}^M A_\nu{}^N \,, & 
\hat \gM_{\mu}{}^\nu & = - g^{\nu  \rho} C_{\mu  \rho} \,,\\
\hat \gM^{\mu \nu} & = g^{\mu \nu} \,, & 
\hat \gM^{\mu}{}_M & = -g^{\mu  \rho} A_{ \rho M}\,, \\
\hat \gM_{\mu M} & = \gM_{MP} A_\mu{}^P + C_{\mu  \rho} g^{ \rho \sigma} A_{\sigma M}\,, & 
\hat \gM_{MN} & = \gM_{MN} + g^{ \rho \sigma} A_{ \rho M} A_{\sigma N} \,,
\end{align} 
with the doubled internal index on $A_\mu{}^M$ now lowered with the $O(d,d)$ structure $\eta_{MN}$, and $C_{\mu \nu} \equiv - B_{\mu \nu} +\frac{1}{2}  A_\mu{}^M A_{\nu M}$.
This interpolates between the fully doubled theory, for $d=D$, and the usual spacetime theory for $d=0$. 
Inserting this into the doubled worldsheet Tseytlin action, splitting the worldsheet coordinates in the same manner and then integrating out $\tilde X_\mu^\prime$, one finds in conformal gauge
\be
\begin{split} 
S_{DWS} &= T \int d^2\sigma \Big( \frac{1}{2} \eta_{MN} D_\tau X^M D_\sigma X^N - \frac{1}{2} \gM_{MN} D_\sigma X^M D_\sigma X^N 
\\ & \qquad \qquad \quad + \frac{1}{2} \eta_{MN} \left( X^{\prime M} A_\nu{}^N \dot{X}^\mu -\dot{X}^M A_\nu{}^N X^{\prime \mu} \right) \\
& \qquad \qquad \quad+ \frac{1}{2} g_{\mu\nu} \left( \dot{X}^\mu \dot{X}^{\nu} - X^{\prime \mu} X^{\prime \nu} \right) + b_{\mu\nu} \dot{X}^\mu X^{\prime \nu}
\Big) \,,
\end{split} 
\label{eq:Tsplit} 
\ee
where
\be
D_\tau X^M \equiv \dot{X}^M + A_\mu{}^M \dot{X}^\mu \quad , \quad 
D_\sigma X^M \equiv {X}^{\prime M} + A_\mu{}^M X^{\prime \mu} \,. 
\ee
This action is supplemented with the Virasoro constraints:
\be
g_{\mu\nu} ( \dot{X}^\mu\dot{X}^\nu + X^{\prime \mu} X^{\prime \nu} ) + \gM_{MN} D_\sigma X^M D_\sigma X^N = 0 \,,
\ee
\be
2 g_{\mu\nu} X^{\prime \mu} \dot{X}^\nu + \eta_{MN}  D_\sigma X^M D_\sigma X^N = 0  \,.
\ee
To identify this with spacetime variables, suppose $\hat g$ and $\hat B$ denote the $D$-dimensional metric and $B$-field, then one has a decomposition
\be
\begin{split}
\hat{g}_{\mu \nu} &  = g_{\mu \nu} + A_\mu{}^p A_\nu{}^q g_{pq} \,, \\ 
\hat{g}_{\mu m} & = A_\mu{}^p g_{pm} \,, \\
\hat{g}_{mn} & = g_{mn}  \,,
\end{split} 
\qquad
\begin{split}
\hat{B}_{\mu \nu} & = B_{\mu \nu} -  A_{[\mu}{}^p A_{\nu] p} + A_\mu{}^p A_\nu{}^q b_{pq}  \,,\\
\hat{B}_{\mu m} & = A_{\mu m} + A_{\mu}{}^p b_{pm}  \,,\\
\hat{B}_{mn} & = b_{mn}  \,,
\end{split}
\ee
with $m,n$ $d$-dimensional spacetime indices, so that we can identify the components of $A_\mu{}^M$ as the $A_\mu{}^m$ and $A_{\mu m}$ appearing here, while $g_{mn}$ and $b_{mn}$ are packaged into the generalised metric $\gM_{MN}$ in the usual fashion. Note that $g_{\mu\nu}$ and $B_{\mu\nu}$ are invariant under T-duality. 
 
\section{Supplementary results: spacetime} 
\label{appst} 

\subsection{The equations of motion of DFT$^{\pm}$}
\label{dfteom}

Here we will vary the DFT action (omitting the prefactor)
\be
S_{DFT^\pm} = \int d^{2D}X e^{-2d} \mathcal{R}^\pm \,,
\ee
where the generalised Ricci scalar is
\be
\begin{split}
\mathcal{R}^\pm & = 4 \gM^{MN}\partial_{M}\partial_{N}d
-\partial_{M}\partial_{N}\gM^{MN} 
-4H^{MN}\partial_{M}d\,\partial_{N}d
+ 4 \partial_M \gM^{MN}  \,\partial_Nd\\
&\pm\frac{1}{8}\gM^{MN}\partial_{M}\gM^{KL}\,
\partial_{N}\gM_{KL}\mp\frac{1}{2}\gM^{MN}\partial_{M}\gM^{KL}
\partial_{K}\gM_{NL}\,.
\end{split}
\label{RDFTboth}
\ee
By definition,
\be
\gM^{MN} \equiv \eta^{MP} \eta^{NQ} \gM_{PQ} \,,
\ee
and we have
\be
\eta^{MN} \gM_{NP} \eta^{PQ} \gM_{QR} = \pm \delta^M_R \,.
\ee
If we vary the above condition, we find that the variation $\delta \gM^{MN}$ must obey
\be
\left(  \delta_M^P \delta_N^Q \pm \gM^P{}_M \gM^Q{}_N \right) \delta \gM^{MN} = 0\,.
\ee
When varying the action, we must be careful to only ever raise or lower indices using $\eta_{MN}$, remembering that $\gM^{MN}$ is not necessarily the inverse of $\gM_{MN}$. The result is, discarding the total derivatives, 
\be
\delta S_{DFT^{\pm}} = \int d^{2D} X e^{-2d} \left( - 2 \delta d \mathcal{R}^{\pm} + \delta \gM^{MN} K_{MN}^\pm \right) \,,
\ee
where
\be
\begin{split}
K_{MN}^\pm & = 
\pm \frac{1}{8} \partial_M \gM^{KL} \partial_N \gM_{KL} 
\mp \frac{1}{2} \partial_{(M|} \gM^{KL} \partial_K \gM_{|N)L}
+ 2 \partial_M \partial_N d 
\\ & 
\pm \left( \partial_P - 2 \partial_P d\right) 
\left( 
- \frac{1}{4} \gM^{PQ} \partial_Q \gM_{MN} 
+ \frac{1}{2} \gM^{PQ} \partial_{(M} \gM_{N)Q} 
+ \frac{1}{2} \gM^{KQ} \eta_{Q(M} \partial_K \gM^{PL} \eta_{N)L} \right) \,.
\end{split}
\ee
The true equation of motion taking into account the constraint on the variation is however given by the following generalised Ricci tensor:
\be
\mathcal{R}_{MN}^\pm = \frac{1}{2} \left( \delta_M{}^P \delta_N{}^Q \mp \gM_M{}^P \gM_N{}^Q \right) K_{PQ}^\pm
\ee
In DFT$^+$ we define the projectors $P_M{}^N = \frac{1}{2} (\delta_M{}^N - \gM_M{}^N)$, $\bar{P}_M{}^N = \frac{1}{2} ( \delta_M{}^N + \gM_M{}^N )$, which in DFT$^-$ should instead be given by $P_M{}^N = \frac{1}{2} (\delta_M{}^N + i \gM_M{}^N)$, $\bar{P}_M{}^N = \frac{1}{2} ( \delta_M{}^N - i\gM_M{}^N )$. Then one sees that in both theories 
\be
\mathcal{R}_{MN}^\pm = \left( P_M{}^P \bar{P}_N{}^Q + \bar{P}_M{}^P P_N{}^Q \right) K_{PQ}^\pm \,.
\ee

\subsection{The generalised Lorentz group of DFT$^-$}
\label{genlor}

\subsubsection{The general structure}

In this subsection, we discuss the form of the local generalised Lorentz transformations of DFT$^-$. These are required to preserve the flattened $O(D,D)$ structure, which we call $\hat{\eta}$, and the flattened generalised metric, $\hat{\gM}$, 
\be
\hat{\eta} = \begin{pmatrix} 0 & I \\ I & 0 \end{pmatrix} 
\quad , \quad 
\hat{\gM} = \begin{pmatrix} - h_{t,s} & 0 \\ 0 & h_{t,s} \end{pmatrix} 
\label{flatstuff}
\ee
where
\be
h_{t,s} = \mathrm{diag}\,( \underbrace{-1,\dots,-1}_t , \underbrace{+1,\dots,+1}_s )
\ee
For convenience, abbreviate $h_{t,s} \equiv h$. 

We denote the group of such transformations by $H$. If we take an element $g \in H$, and write it as
\be
g = \begin{pmatrix} A & B \\ C&  D \end{pmatrix} 
\ee
then the conditions $g^T \hat{\eta} g = \hat{\eta}$ and $g^T \hat{\gM} g = \hat{\gM}$ imply respectively that
\be
A^T C + C^T A = 0 \quad ,\quad
B^T D + D^T B = 0 \quad ,\quad
A^T D + C^T B = 1 \,,
\ee
\be
 - A^T h A + C^T h C = - h \quad,\quad
- B^T h B + D^T h D = h \quad,\quad
- A^T h B + C^T h D = 0  \,.
\ee
It is more convenient to write $g = I + X$ and find the conditions on $X$ to be an element of the Lie algebra of $H$, which we call $L(H)$. The matrix $X$ must satisfy
\be
X = - \hat{\eta}^{-1} X^T \hat{\eta} 
\quad , \quad
X = - \hat{\gM}^{-1} X^T \hat{\gM}\,,
\ee
and the general form of $X$ obeying these conditions is parametrised in terms of two $d \times d$ matrices $a$ and $b$ as
\be
X = \begin{pmatrix} a & b \\ - h bh & h a h \end{pmatrix} 
\label{X}
\ee
with the constraints
\be
b^T = - b \quad , \quad a^T = - h a h \,.
\ee
The latter is just the condition that $a \in so(t,s)$. As a result, the dimension of $L(H)$ is $d(d+1)$. 

Let us write the Lie algebra elements involving solely $a$ as
\be
X_a = \begin{pmatrix} a & 0 \\ 0 & hah \end{pmatrix} 
\quad,\quad a^T = - hah \,.
\ee
These exponentiate to an $SO(t,s)$ subgroup of $H$, of the form
\be
g_A = \begin{pmatrix} A & 0 \\ 0 & A^{-T} \end{pmatrix} 
\quad,\quad A = e^a \in SO(t,s) \,.
\ee
Meanwhile, we also write 
\be
\tilde X_b = \begin{pmatrix} 0 & b \\ - hbh & 0 \end{pmatrix} 
\quad,\quad b^T = - b \,.
\ee
We can compute the commutation relations in terms of these elements:
\be
[X_a,X_{a^\prime} ] = X_{[a,a^\prime]} 
\quad,\quad
[ X_a,\tilde X_b ] = \tilde X_{ab+ba^T} 
\quad,\quad
[
\tilde X_b, \tilde X_{b^\prime} 
] 
= X_{ - b h b^\prime h + b^\prime h bh } \,.
\label{commX}
\ee
Now, looking at the definitions \eqref{flatstuff} we note that the transformations which permute the diagonal entries of $\hat{\gM}$ are numerically equal to the matrix form of Buscher transformations. Hence, they leave $\hat{\eta}$ unchanged. The form of $H$ will not be changed by such permutations, as they are just a reordering of the coordinates. We can therefore specialise to the simplest case of $t=0$ and $s=D$, so that $h \equiv h_{0,D} = I$. 

In this case, the matrices $a$ and $b$ appearing in the Lie algebra element \eqref{X} are both antisymmetric. 
Each of these give $\frac{1}{2} D(D-1)$ generators of $O(D)$. Let us denote the generators of $O(D)$ by $e_A$, such that $[e_A,e_B]= f_{AB}{}^C e_C$ (the index $A$ is an algebra index here, running from $1$ to $\frac{1}{2}d(d-1)$). Then we can take the following generators for $L(H)$:
\be
E_A = \begin{pmatrix} e_A & 0 \\ 0 & e_A \end{pmatrix} \quad ,\quad
\tilde E_A = \begin{pmatrix} 0 & e_A \\ - e_A & 0 \end{pmatrix} \,,
\ee
which obey the commutation relations
\be
[ E_A , E_B ] = f_{AB}{}^C E_C \quad , \quad
[ E_A , \tilde E_B ] = f_{AB}{}^C  \tilde E_C \quad ,\quad
[ \tilde E_A , \tilde E_B ] = - f_{AB}{}^C E_C \,.
\label{exoticodod}
\ee
The $E_A$ give the spacetime Lorentz group, which here is $O(D)$. The above algebra also contains all possible $O(t,s)$ subalgebras, by choosing different sets of the $E_A$ and $\tilde E_A$. 

For instance, in $D=3$, one sees that $E_1,E_2,E_3$ generate the $O(3)$ group corresponding to the Euclidean space with $t=0$ and $s=3$. If we exchange the $1$ direction for a timelike direction, then one finds that $E_1, \tilde E_2$ and $-\tilde E_3$ generate the corresponding $O(1,2)$ Lorentz group. 

\subsubsection{Relationship to $O(D) \times O(D)$}

We can compare the above commutation relations with the ones that arise for the conventional generalised metric (with Euclidean signature). Taking $\hat{H}$ to be the identity, we find that the generators $E_A$ are unchanged but that there is no minus sign in the bottom left block of $\tilde E_A$. As a result, we have commutation relations
\be
[ E_A , E_B ] = f_{AB}{}^C E_C \quad 
[ E_A , \tilde E_B ] = f_{AB}{}^C  \tilde E_C \quad 
[ \tilde E_A , \tilde E_B ] = f_{AB}{}^C E_C 
\label{ordinaryodod}
\ee
We have not made any assumptions about reality of coefficients, so that this can be taken to correspond to the Lie algebra of $O(D;\mathbb{C})\times O(D;\mathbb{C})$, generated by $t_A^\pm = \frac{1}{2} ( E_A\pm \tilde E_A)$. 

Restricting to solely real coefficients gives the compact real form, $O(D;\mathbb{R}) \times O(D;\mathbb{R})$ which we normally refer to just as $O(D)\times O(D)$. One can choose the coefficients of some of the generators to be purely imaginary to obtain alternative real forms. Doing this for $p(D-p)$ of the $t_A^+$ and the same $p(D-p)$ of the $h_A^-$ (so $p$ of the $E_A$ and the same $p$ $\tilde E_A$) leads to the split real forms $O(p,D-p) \times O(p,D-p)$.

However, to obtain the algebra \eqref{exoticodod} we need the replacement $\tilde E_A \rightarrow \pm i \tilde E_A$. As a result, \eqref{exoticodod} is simply a different real form of \eqref{ordinaryodod}. However, generically there is no guarantee that \eqref{exoticodod} is of the form $O(p,D-p) \times O(q,D-q)$. Indeed, by counting the number of compact and non-compact generators, one finds for instance that this is impossible in $D=3,6,7$.\footnote{In e.g. $D=10$ the only possibility is $O(4,6) \times O(3,7)$ (up to swapping the time and space directions in each factor), in $D=26$ it is $O(16,10)\times O(15,11)$. However, we have not verified if these actually correspond to the algebra obtained. In $D=3$, in fact, the algebra \eqref{exoticodod} is isomorphic to that of $\mathrm{sl}(2;\mathbb{C})$.}

Let us write a generic element of the complex algebra as $X = \alpha_A E_A + \tilde \alpha_A \tilde E_A = ( \alpha_A + \tilde \alpha_A ) t_A^+ + ( \alpha_A - \tilde\alpha_A ) t_A^-$, with $\alpha_A,\tilde\alpha_A\in \mathbb{C}$. We restrict to real $\alpha_A \rightarrow a_A$ and imaginary $\tilde \alpha_A \rightarrow \tilde b_A$, for $a_A, \tilde b_A \in \mathbb{R}$. Then we see that if we continue to express $X \in H$ in terms of the $O(D) \times O(D)$ generators $t_A^\pm$, we have $X = ( a_A + i \tilde b_A ) t_i^+ + ( a_A - i \tilde b_A ) t_A^-$, showing that this restriction corresponds to taking an $O(D)$ subgroup along with its complex conjugate. 

\subsubsection{Two examples with $D=2$}

As an illustration of how different spacetime Lorentz groups are contained within $H$, let us look at the very simple example of $D=2$. We start with the $t=0,s=2$ case. Defining
\be
\epsilon = \begin{pmatrix} 0 & 1 \\ - 1 & 0 \end{pmatrix} \,,
\ee
then the two generators of $L(H)$ can be taken to be
\be
E = \begin{pmatrix} \epsilon & 0 \\ 0 & \epsilon \end{pmatrix} \quad , \quad 
\tilde E = \begin{pmatrix} 0 & \epsilon \\ \epsilon & 0 \end{pmatrix} \,.
\ee
These commute, and exponentiate to give an $SO(2)$ and $SO(1,1)$ subgroup, respectively, consisting of matrices of the form
\be
g_\alpha = \begin{pmatrix} A & 0 \\ 0 & A \end{pmatrix} 
\quad,\quad A = \begin{pmatrix} \cos \alpha & \sin \alpha \\ -\sin \alpha & \cos \alpha \end{pmatrix} \,,
\ee
and
\be
g_\beta = \begin{pmatrix} 
\cosh \beta I & \sinh \beta \epsilon \\
- \sinh \beta \epsilon & \cosh \beta I 
\end{pmatrix} \,.
\ee
Note that $g_\beta g_{\beta^\prime} = g_{\beta+\beta^\prime}$, $g_\alpha g_\beta = g_\beta g_\alpha$. We conclude that $H = O(1,1) \times O(2)$.  

Now instead let $t=1$ and $s=1$, so now $h= \mathrm{diag}(-1,1)$. Returning to \eqref{X}, we find that the generators in this case can be taken to be
\be
E^\prime = \begin{pmatrix} - \eta & 0 \\ 0 &  \eta \end{pmatrix} 
\quad , \quad 
\tilde E^\prime = \begin{pmatrix} 0 & \epsilon \\ \epsilon & 0 \end{pmatrix} \quad , \quad 
\eta \equiv \begin{pmatrix} 0 & 1 \\ 1 & 0 \end{pmatrix} \,.
\ee
The flat generalised metric $\gM^\prime$ and generators are related to those of the $t=0,s=2$ case by
\be
\gM^\prime = P^T \gM P \quad , \quad 
E^\prime = P^T \tilde E P \quad , \quad 
\tilde E^\prime = P^T E P 
\ee
for the permutation matrix
\be
\footnotesize
P = \begin{pmatrix} 
0 & 0 & 1 & 0 \\ 
0 & 1 & 0 & 0 \\
1 & 0 & 0 & 0 \\
0 & 0 & 0 & 1 
\end{pmatrix} \,.
\ee
Exponentiating the $E^\prime$ generator yields
\be
g^\prime_{\beta}  = \begin{pmatrix} A & 0 \\ 0 & A^{-1} \end{pmatrix} 
\quad , \quad A = \begin{pmatrix} \cosh \tilde \beta & -\sinh \tilde \beta \\- \sinh \tilde \beta & \cosh \tilde \beta \end{pmatrix} \,,
\ee
giving the $SO(1,1)$ subgroup, while exponentiating the $\tilde E^\prime$ generator gives
\be
g^\prime_{\alpha} = \begin{pmatrix} \cos \tilde\alpha I & \sin \tilde \alpha \epsilon \\ \sin \tilde \alpha \epsilon & \cos \tilde\alpha I \end{pmatrix} \,,
\ee
giving the $SO(2)$ subgroup. We see that again $H = O(1,1) \times O(2)$ is fixed, and the interpretation of which subgroup is the spacetime Lorentz group differs. 

\subsection{Generalised metrics for strings with momentum and winding} 
\label{dstube} 

In this section we record the possibly useful expression for the generalised metric of some interesting configurations involving vibrating strings carrying winding and momentum. We solely consider the usual DFT$^+$ theory. 
Given a metric and $B$-field of the form
\be
\begin{split} 
ds^2 & = H^{-1} \left( (f-2) dt^2 + 2(f-1) dt dz + f dz^2 + 2 A_i dx^i ( dt+ dz) \right) + \delta_{ij} dx^i dx^j \,,\\
B & = (H^{-1} - 1) dt \wedge dz + H^{-1} A_i (dt+ dz) \wedge dx^i\,,
\end{split}
\ee
then the generalised metric for $X^M = ( t,z,\tilde t, \tilde z, i, \tilde i)$ is \footnotesize
\be
\gM_{MN} = 
\begin{pmatrix} 
(f-2)(2-H) +A^2 & (f-1)(2-H)+A^2 & (1-H)(f-1) + A^2 & (1-H)(2-f) - A^2 & A_j & A^j \\
(f-1)(2-H) + A^2 & f(2-H) + A^2 & - f(H-1) +A^2 & (H-1)(f-1) - A^2 & A_j & A^j \\
(1-H)(f-1) + A^2 & -f(H-1)+A^2 & -fH+A^2 & H(f-1)-A^2 &A_j & A^j \\ 
(1-H)(2-f) - A^2 & (H-1)(f-1)-A^2 & H(f-1)-A^2 & H(2-f)+A^2 & -A_j & -A^j \\
 A_i & A_i & A_i & -A_i & \delta_{ij} &0  \\
 A^i & A^i & A^i &  - A^i & 0 & \delta^{ij} 
\end{pmatrix} \,.
\label{gengenmet}
\ee
\normalsize 
A Buscher duality on the $z$ direction has the effect of interchanging $f$ and $H$. 

The vibrating string solution \eqref{vibstr} corresponds to taking
\be
H = 1 + \frac{Q}{|\vec{x} - \vec{F}(v)|^6} \quad , \quad
f = 1+ \frac{ Q \dot{F}^2 }{|\vec{x} - \vec{F}(v)|^6} \quad,\quad
A_i = \frac{-Q \dot{F}_i}{|\vec{x} - \vec{F}(v)|^6} \,.
\label{vibstrHfa}
\ee
Superimposing many ``strands'' of an oscillating string, taking the limit of large quanta of momenta and winding and also smearing over some number of the transverse coordinates leads to supertube configurations which can be dualised (using both T- and S-dualities) to other interesting systems (a useful account, in the context of a review relating to fuzzballs and the D1-D5 systems, is \cite{Mathur:2005zp}). If we smear over five transverse coordinates, then the solution specified by the following configuration
\be
H = 1 + \frac{Q}{L} \int_0^L \frac{dv}{|\vec{x} - \vec{F}(v)|} \quad ,\quad
f = 1 + \frac{Q}{L} \int_0^L \frac{dv \dot{F}^2}{|\vec{x} - \vec{F}(v)|} \quad ,\quad
A_i = - \frac{Q}{L}  \int_0^L \frac{dv \dot{F}_i}{|\vec{x} - \vec{F}(v)|} \,.
\label{stube}
\ee
(here we only depend on three transverse coordinates $\vec{x}$, so that $i=1,2,3$) can be related by duality to a supertube configuration D4+D4 $\rightarrow 5_2^2 + $p where two D4 branes puff-up to produce an exotic $5_2^2$ brane dipole \cite{deBoer:2012ma}. The generalised metric \eqref{gengenmet} and configuration \eqref{stube} represent the F1+P $\rightarrow$ f1 + p supertube in DFT. 

Note that it does not seem obvious how one could directly carry out this superposition starting from the generalised metric: observe for instance that the component $\gM_{t}{}^z = (1-H)(f-1) + A^2$ vanishes for the configuration \eqref{vibstrHfa}, but is non-zero for the solution specified by \eqref{stube}.

%\bibliography{Bib}
%\bibliographystyle{JHEP}

\providecommand{\href}[2]{#2}\begingroup\raggedright\endgroup

\end{document}